\documentclass[a4paper,12pt]{article}

\pdfoutput=1

\usepackage[paper=letterpaper,margin=1in]{geometry}

\usepackage{amsmath,amssymb,amsfonts,epsfig,cite,setspace,bigstrut,longtable,array,breqn}

\begin{document}


\vskip 0.25in

\newcommand{\todo}[1]{{\bf ?????!!!! #1 ?????!!!!}\marginpar{$\Longleftarrow$}}
\newcommand{\fref}[1]{Figure~\ref{#1}}
\newcommand{\tref}[1]{Table~\ref{#1}}
\newcommand{\sref}[1]{\S~\ref{#1}}
\newcommand{\nn}{\nonumber}
\newcommand{\tr}{\mathop{\rm Tr}}
\newcommand{\comment}[1]{}

\newcommand{\cM}{{\cal M}}
\newcommand{\cW}{{\cal W}}
\newcommand{\cN}{{\cal N}}
\newcommand{\cH}{{\cal H}}
\newcommand{\cK}{{\cal K}}
\newcommand{\cZ}{{\cal Z}}
\newcommand{\cO}{{\cal O}}
\newcommand{\cB}{{\cal B}}
\newcommand{\cC}{{\cal C}}
\newcommand{\cD}{{\cal D}}
\newcommand{\cE}{{\cal E}}
\newcommand{\cF}{{\cal F}}
\newcommand{\cX}{{\cal X}}
\newcommand{\IA}{\mathbb{A}}
\newcommand{\IP}{\mathbb{P}}
\newcommand{\IQ}{\mathbb{Q}}
\newcommand{\IH}{\mathbb{H}}
\newcommand{\IR}{\mathbb{R}}
\newcommand{\IC}{\mathbb{C}}
\newcommand{\IF}{\mathbb{F}}
\newcommand{\IV}{\mathbb{V}}
\newcommand{\II}{\mathbb{I}}
\newcommand{\IZ}{\mathbb{Z}}
\newcommand{\re}{{\rm Re}}
\newcommand{\im}{{\rm Im}}
\newcommand{\sym}{{\rm Sym}}

\newcommand{\tmat}[1]{{\tiny \left(\begin{matrix} #1 \end{matrix}\right)}}
\newcommand{\mat}[1]{\left(\begin{matrix} #1 \end{matrix}\right)}
\newcommand{\diff}[2]{\frac{\partial #1}{\partial #2}}
\newcommand{\gen}[1]{\langle #1 \rangle}
\newcommand{\ket}[1]{| #1 \rangle}
\newcommand{\jacobi}[2]{\left(\frac{#1}{#2}\right)}

\newcommand{\drawsquare}[2]{\hbox{%
\rule{#2pt}{#1pt}\hskip-#2pt
\rule{#1pt}{#2pt}\hskip-#1pt
\rule[#1pt]{#1pt}{#2pt}}\rule[#1pt]{#2pt}{#2pt}\hskip-#2pt
\rule{#2pt}{#1pt}}
\newcommand{\fund}{\raisebox{-.5pt}{\drawsquare{6.5}{0.4}}}
\newcommand{\antifund}{\overline{\fund}}

\newtheorem{theorem}{\bf THEOREM}
\def\thetheorem{\thesection.\arabic{theorem}}
\newtheorem{proposition}{\bf PROPOSITION}
\def\thetheorem{\thesection.\arabic{proposition}}
\newtheorem{observation}{\bf OBSERVATION}
\def\thetheorem{\thesection.\arabic{observation}}

\def\theequation{\thesection.\arabic{equation}}
\newcommand{\setall}{\setcounter{equation}{0}
        \setcounter{theorem}{0}}
\newcommand{\setequation}{\setcounter{equation}{0}}
\renewcommand{\thefootnote}{\fnsymbol{footnote}}

~\\
\vskip 1cm

\begin{center}
{\Large \bf Eta Products, BPS States and K3 Surfaces}
\end{center}
\medskip

\vspace{.4cm}
\centerline{
{\large Yang-Hui He}$^1$ \&
{\large John McKay}$^2$
}
\vspace*{3.0ex}

\begin{center}
{\it
{\small
{${}^{1}$ 
Department of Mathematics, City University, London, EC1V 0HB, UK and \\
School of Physics, NanKai University, Tianjin, 300071, P.R.~China and \\
Merton College, University of Oxford, OX14JD, UK\\
\qquad hey@maths.ox.ac.uk\\
}
\vspace*{1.5ex}
{${}^{2}$ 
Department of Mathematics and Statistics,\\
Concordia University, 1455 de Maisonneuve Blvd.~West,\\
Montreal, Quebec, H3G 1M8, Canada\\
\qquad mckay@encs.concordia.ca
}
}}
\end{center}

\vspace*{4.0ex}
\centerline{\textbf{Abstract}} \bigskip
Inspired by the multiplicative nature of the Ramanujan modular discriminant, $\Delta$, we consider physical realizations of certain multiplicative products over the Dedekind eta-function in two parallel directions: the generating function of BPS states in certain heterotic orbifolds and elliptic K3 surfaces associated to congruence subgroups of the modular group.
We show that they are, after string duality to type II, the same K3 surfaces admitting Nikulin automorphisms.
In due course, we will present identities arising from q-expansions as well as relations to the sporadic Mathieu group $M_{24}$.
\newpage

\tableofcontents

\section{Introduction and Motivation}
\setcounter{footnote}{0}
On the virtues of 24 there has been much written.
The ever expanding tapestry of correspondences, intricate and beautiful, continues to be woven in many new directions.
Of the multitude of the stories surrounding this mysterious number we will isolate two strands of thought, both rich in mathematics and in physics, touching upon such diverse subjects as K3 surfaces, multiplicative functions, BPS counting and modular subgroups.

We begin with the standard fact that the Dedekind eta-function $\eta(q) = q^{\frac{1}{24}} \prod\limits_{n=1}^\infty (1- q^n)$ has a prefactor of $q^{\frac{1}{24}}$ which is crucial to its quasi-modularity (cf.~Appendix \ref{ap:eta-modular}).
It is also well-known, dating at least to Euler, that once removing this factor, the reciprocal is the generating function for the partition of positive integers.
This fact was exploited in the computation of oscillator modes in string theory. Interestingly, for the bosonic string, which is critical in 26 dimensions, the physical degrees of freedom, upon quantization in the light-cone, is a counting governed by $q \eta(q)^{-24}$, i.e., by the free partition of integers in 24 independent directions (spatial dimensions).
What is perhaps less appreciated is the fact that the reciprocal generating function, $\eta(q)^{-24}$, enjoys extraordinary properties: the expansion coefficients are the values of the famous Ramanujan tau-function, and are of a remarkable multiplicative nature.

On this latter point, quite independent of physical interpretation, the natural and important question of what other products of $\eta$-functions, viz., functions of the form $\prod_i\eta(q^{a_i})^{b_i}$ for some finite set of positive integers $a_i,b_i$, was addressed in \cite{DKM}.
These turn out to correspond to exactly 30 partitions of 24 and are all modular forms of appropriate weight, level and character.

Thus, the question which instantly emerges is whether it is possible to find physical systems whose partition functions are exactly these elegant products.
Remarkably, at this list did the authors of \cite{Govindarajan:2009qt} arrive when considering the counting of electrically charged, 1/2-BPS states in the $\cN=4$ supersymmetric CHL orbifolds of the heterotic string on the six-torus.
What is particularly fascinating for our present purposes is that upon string duality the situation is equivalent to the compactification of type IIB string theory on the product of a two-torus with a K3 surface of a specific type, viz., one which admits certain Nikulin involutions.
Such K3 are quite special \cite{nikulin} and there are 14 in type, having automorphisms which are various Abelian groups of fairly low order.

In a parallel vein, there is an equally valuable partition problem of 24 involving K3 surfaces.
This is the list of semi-stable extremal elliptic fibrations and constituted the classification of \cite{MP}, numbering a total of 112.
Such K3 surfaces 
have maximal Picard number and, more relevant for our present discourse, of having elliptic j-invariants which are Belyi maps from $\IP^1$ to $\IP^1$ and thus are associated to dessins d'enfants \cite{BM,He:2012jn}.
A special class has been distilled in order to study congruence subgroups of the modular groups, in relation to Seiberg-Witten curves of certain $\cN=2$ gauge theories in four space-time dimensions \cite{He:2012kw}.
And thus we are brought to the final list of our dramatis personae, which are torsion-free genus zero congruence subgroups of the modular group; such have been classified in \cite{mckaysebbar,sebbar,classSebbar}, tallying 33 in total.
The Schreier coset graphs of these are also trivalent and clean (all, say,  white nodes are valency two) dessins d'enfants.
At the intersection of the extremal 112 K3 list and the 33 congruence subgroup list lie 9 distinguished K3 surfaces which are modular elliptic.

The above information we shall introduce in detail in \S\ref{s:eta} and \S\ref{s:K3}, discussing, as we encounter the objects of our concern, the relevant quantities in our parallel context.
In due course, we shall show by explicit computation, that the extremal K3 surfaces from the Nikulin/CHL side coincides with the congruence/extremal side, at least for semi-stable models of the former.
In fact, we will see that one can go beyond extremality and establish correspondence between the Nikulin list and the eta-product list for all partitions of 24 not with at least 6 parts.
Emboldened, having touched upon the fact that our eta-products are not only multiplicative, but are also modular forms, it is irresistible not to enter the realm of elliptic curves as guided by Shimura-Taniyama-Wiles.
The more general situation of which eta-quotients - i.e., allowing our aforementioned integers $b_i$ to be negative as well - are weight two modular forms was investigated in \cite{ono}.
We will focus on our eta-products with four partitions which produce weight 2 modular forms and study the corresponding elliptic curves explicitly.
Of equal importance is how the dessins relate to so-called ''Mathieu Moonshine''. The fundamental eta-product, namely the Ramanujan-tau function corresponding to the partition $1^{24}$, already encodes the irreducible representations of the sporadic group $M_{24}$. We will present the associated dessins in \S\ref{s:moonshine}.

As a parting digression, we will take an alternative physical interpretations from the perspectives of the Plethystic Programme in \S\ref{s:pleth}, which is a method of extracting underlying geometries from the generating function of half-BPS states by computing certain Hilbert series via an inverse Euler transform.
Finally, in \S\ref{s:conc}, we conclude with prospects and outlook.

The interested reader might find the Appendix entertaining; therein we will take a rapid excursion on a multitude of expansions and identities, mostly rudimentary but some less so. 
In Appendix A we collect some standard facts on various modular and partitioning properties of the Dedekind eta-function.
In Appendix B, we will exploit the relation of the j-invariant to the eta-function, and thence the Euler phi-function, to express the former in terms of the partition of integers, as well as these partitions in terms of divisor functions. Moreover, we will amuse ourselves with q-expansions of various $n$-th roots of the j-invariants for $n$ being a divisor of 24, including the example of the cube-root, which is known to encode the representations of the $E_8$ Lie group \cite{kac1,kac2}.

\subsection{Nomenclature}
Before turning to the full exposition of our tale, since we shall alight upon a variety of objects, for clarity we will adhere to the following standard notation which we summarize here.
\begin{itemize}
\item The upper-half plane $\{z : \im(z) > 0\}$ is denoted as $\cH$, with coordinate $z$ and nome $q$:
\begin{equation}
z \in \cH \ , \quad q = \exp( 2 \pi i z) \ .
\end{equation}

\item
The Euler phi-function (we will use this one rather than the reciprocal)
\begin{equation}\label{phi}
\varphi(q) = \prod\limits_{n=1}^\infty (1- q^n)^{-1} = \sum_{k=0}^\infty
\pi_k q^k
\end{equation}
is the formal generating function for the partition $\pi_k$ of integers $k \in \IZ_{\ge 0}$.

\item
The Dedekind eta-function is related to $\varphi(q)$ as
\begin{equation}
\eta(z) = q^{\frac{1}{24}} \prod\limits_{n=1}^\infty (1- q^n) = 
q^{\frac{1}{24}} \varphi(q)^{-1}
\end{equation}

\item The Jacobi theta functions are defined with the following conventions
\begin{align}
\nn
\theta_1(q,y) = i \sum\limits_{n=-\infty}^\infty (-1)^n q^{\frac{(n-\frac12)^2}{2}} y^{n-\frac12} \ , \quad 
\theta_2(q,y) = \sum\limits_{n=-\infty}^\infty q^{\frac{(n-\frac12)^2}{2}} y^{n-\frac12} \ , \\ \label{theta}
\theta_3(q,y) = \sum\limits_{n=-\infty}^\infty q^{\frac{n^2}{2}} y^{n} \ , \quad 
\theta_4(q,y) = \sum\limits_{n=-\infty}^\infty (-1)^n q^{\frac{n^2}{2}} y^{n} \ , 
\end{align}
with $q = \exp(2\pi i z)$ and $y = \exp(2\pi i \tilde{z})$.
Moreover, the single argument case is understood to be $\theta_i(q) := \theta_i(q, 1)$ for all $i=1,2,3,4$.

\item
The modular discriminant $\Delta$ and Ramanujan tau-function $\tau(n)$ are related to eta by:
\begin{equation}
\Delta(z) = \eta(z)^{24} :=  q\prod\limits_{n=1}^\infty (1- q^n)^{24} = 
\sum\limits_{n =1}^\infty \tau(n) q^n \ .
\end{equation}
In term of the Weierstra\ss\ form of an elliptic curve
\begin{equation}\label{weier}
y^2 = 4x^3 -g_2 \ x - g_3 \ ,
\end{equation}
the discriminant is $\Delta = g_2^3 - 27 g_3^2$ and the modular j-invariant is
$j = 1728 \frac{g_2^3}{\Delta}$.
We will use upper case $J$ to refer to the J-invariant, which is $j$ without the $12^3=1728$ prefactor.

\item A level $N$ weight $k$ modular form with character $\chi$ is a holomorphic function $f(z)$ which transforms under the congruence group $\Gamma_0(N) \subset SL(2;\IZ)$ as 
\begin{equation}
f(\frac{a z + b}{c z + d}) = (c z + d )^k \chi^k f(z) \ ,  \quad
\Gamma_0(N) := \left\{ \left. \left( \begin{matrix}
a & b \\ c & d
\end{matrix} \right) \in SL(2;\IZ)  \, \right| \, c \equiv 0 \bmod N \right\} / \{\pm I \}
\end{equation}

\item The Jacobi symbol, for $a \in \IZ$ and odd $n \in \IZ_{> 0}$ with prime factorization $n = p_1^{m_1}  p_2^{m_2} \cdots p_k^{m_k}$, is
\begin{equation}
\jacobi{a}{n} = \jacobi{a}{p_1}^{m_1} \jacobi{a}{p_2}^{m_2} \cdots 
\jacobi{a}{p_k}^{m_k} \ ,
\end{equation}
where for primes in the ``denominator'', we have the Legendre symbol
\begin{equation}
\jacobi{a}{p} = \left\{
\begin{array}{rcl}
0 & \mbox{ if } & a \equiv 0 \bmod p \ ,\\
1 & \mbox{ if } & a \not\equiv 0 \bmod p \mbox{ and } \exists x \in \IZ, a \equiv x^2 \bmod p\ ,\\
-1 & \mbox{ if } & \not\!\exists \mbox{ such } x
\end{array}
\right.
\end{equation}
\end{itemize}

~\\
~\\

\section{Eta Products and Partition Functions}\label{s:eta}
As advertized in the Introduction, we now turn to the details of how products of eta functions as well as their reciprocals enumerate interesting problems, especially in the context of string theory.
We begin with the classic toy example of the bosonic string before turning to a class of partition functions for BPS states in compactifications on certain K3 surfaces.

\subsection{Bosonic String Oscillators}\label{s:boson}\setall
The physical states of the bosonic string \cite{GSW} is given by $\alpha^i_n \ket{0}$, which has mass $\alpha' M^2 = n-1$ and $i=1, \ldots, 24$ refer to the 24 
directions transverse to the light-cone within the famous 26 dimensions, whereby representing the physical oscillations.
Therefore, using the number operator $N := \sum\limits_{n=1}^\infty \alpha_{-n} \cdot \alpha_n = \sum\limits_{n=1}^\infty \sum\limits_{i=1}^{24} \alpha_{-n}^i \alpha_n^i$, the generating function $G(q)$ for the number of states $d_n$  
\begin{equation}\label{Gbosonic}
G(q) = \tr q^{\ \sum\limits_{n=1}^\infty \alpha_{-n} \cdot \alpha_n} = \sum\limits_{n=0}^\infty d_n q^n = \varphi(q)^{24} = q \eta(q)^{-24} = q \Delta(q)^{-1} \ .
\end{equation}

Therefore, whereas $G(q)$ is the generating function for counting the physical states, its reciprocal is the modular discriminant. More precisely, its series expansion gives the Ramanujan $\tau$-function.
\begin{equation}\label{eta24}
qG(q)^{-1} = \sum\limits_{n =1}^\infty \tau(n) q^{n} \ .
\end{equation}
Crucially, the tau-function \cite{serre} is {\bf weakly multiplicative}:
\begin{equation}
\tau(m \ n) = \tau(m) \tau(n) \ , \qquad \mbox{ if } \gcd(m,n) = 1 \ . 
\end{equation}
We need to emphasize that the r\^ole of 24 is essential here, other powers of the Dedekind eta-function would not have this multiplicativity.
For one thing, the pre-factor of $q^{-\frac{1}{24}}$ is a very deep property of the said function, especially in light of its transformations under the modular group \cite{serre,siegel}.
In Appendix \ref{ap:eta-modular}, we will summarize the origin of this 24.

For multiplicative functions, the natural course of action is to take the Dirichlet transform; here we produce the tau-Dirichlet series:
\begin{equation}
T(s) = \sum\limits_{n=1}^\infty \tau(n) n^{-s} \ .
\end{equation}
The zeros of $T(s)$, like those of the Riemann zeta-function, are well-known to have fascinating behaviour; to this point we will return in \S\ref{s:modular}.

\subsection{Eta Products}
The question of whether other combinations of Dedekind eta functions should be multiplicative was posed and answered in \cite{DKM} (q.v.~also \cite{GS} and generalizations to quotients of eta-functions \cite{quotient,ono}; note that the cases of weight 2 and relation to elliptic curves are of particular interest due to the works of Taniyama-Shimura-Wiles).
In particular, products of the form
\begin{equation}\label{etaprod}
F(z) = [n_1, n_2, \ldots, n_t] := \prod\limits_{i=1}^t \eta(n_i z)
\end{equation}
were considered.
The notation $[n_1, \ldots, n_t]$ is commonly called a {\bf frame shape} (or disjoint cycle shape) and $t$, the {\it cycle length}.
Immediately, multiplicativity implies that the frame shape is a partition of 24 and that $a_1$, the coefficient of the linear term in the, q-expansion, is unity.
Interestingly, the motivation for considering such products was in relation to the cycles in the permutation representation of the Mathieu group $M_{24}$; indeed, monstrous behaviour for $M_{24}$ has recently become an active subject
(cf.~\cite{Gannon:2012ck} for a up-to-date review as well as the references therein).

In all, of the $\pi(24) = 1575$ partitions of 24, there are only 30 corresponding eta-products which give multiplicative series expansions.
These are summarized in Table \ref{t:the30}.
We have organized the eta-products according to weight $k$ and level $N$ of which the product is a modular form with character $\chi$ which are either 1 or some Jacobi symbol.
In general, it was shown that under ${\tiny \left( \begin{matrix}
a & b \\ c & d \end{matrix} \right)} \in \Gamma_0(N)$,
\begin{equation}\label{modulartrans}
F(\frac{a z + b}{c z + d}) = (c z + d )^{k} \chi^k F(z) \ , \qquad 
\chi = 
\left\{
\begin{array}{lcl}
(-1)^{\frac{d-1}{2}} \jacobi{N}{d} & \ ,& d \mbox{ odd }\\
\jacobi{d}{N} & \ ,& d \mbox{ even } \ ,
\end{array}
\right.
\end{equation}
and $t = 2k$ is the number of the parts (cycle length). Note that for two cases $k$ is a half-integer.
Note that the fact the Jacobi symbol is only defined for odd ``denominator'' is not a problem here since when $d$ is even, $ad-bc=1$ whilst $c \equiv 0 \bmod N$
implies that $N$ must be odd.

\begin{table}[h!!!]
\begin{center}
\begin{tabular}{ccccc}
\begin{tabular}{|l|l|c|c|} \hline
$k$ & $N$ & eta-product & $\chi$  \\  \hline  \hline
12 & 1 & $[1^{24}]$ & 1 \\  \hline

8 & 2 & $[2^8, 1^8]$ & 1 \\  \hline

6 & 3 & $[3^6,1^6]$ & 1 \\ 
  & 4 & $[2^{12}]$ & 1 \\  \hline

5 & 4 & $[4^4,2^2,1^4]$ & $\jacobi{-1}{d}$ \\  \hline

4 & 6 & $[6^2,3^2,2^2,1^2]$ & 1 \\
  & 5 & $[5^4,1^4]$ & 1 \\
  & 8 & $[4^4,2^4]$ & 1 \\
  & 9 & $[3^8]$ & 1 \\  \hline
3 & 8 & $[8^2,4,2,1^2]$ & $\jacobi{-2}{d}$ \\
  & 7 & $[7^3,1^3]$ & $\jacobi{-7}{d}$ \\
  & 12 & $[6^3,2^3]$ & $\jacobi{-3}{d}$ \\
  & 16 & $[4^6]$ & $\jacobi{-1}{d}$ \\  \hline
\end{tabular}
& \qquad &
\begin{tabular}{|l|l|c|c|} \hline
$k$ & $N$ & eta-product & $\chi$  \\  \hline \hline
2 & 15 & $[15,5,3,1]$ & 1 \\
  & 14 & $[14,7,2,1]$ & 1 \\
  & 24 & $[12,6,4,2]$ & 1 \\
  & 11 & $[11^2,1^2]$ & 1 \\
  & 20 & $[10^2,2^2]$ & 1 \\
  & 27 & $[9^2,3^2]$ & 1 \\
  & 32 & $[8^2,4^2]$ & 1 \\
  & 36 & $[6^4]$ & 1 \\
 \hline

1 & 23 & $[23,1]$ & $\jacobi{-23}{d}$ \\
  & 44 & $[22,2]$ & $\jacobi{-11}{d}$ \\
  & 63 & $[21,3]$ & $\jacobi{-7}{d}$ \\
  & 80 & $[20,4]$ & $\jacobi{-20}{d}$ \\ 
  & 108 & $[18,6]$ & $\jacobi{-3}{d}$ \\ 
  & 128 & $[16,8]$ & $\jacobi{-2}{d}$ \\ 
  & 144 & $[12^2]$ & $\jacobi{-1}{d}$ \\ 
 \hline
\end{tabular}
& \qquad &
\begin{tabular}{|l|c|} \hline
$k$ & eta-product  \\  \hline \hline
``$\frac32$'' & $[8^3]$ \\ \hline
``$\frac12$'' & $[24]$ \\ \hline
\end{tabular}
\end{tabular}
\caption{{\sf
The 30 multiplicative eta-products, organized by weight $k$ and level $N$
for the congruence group $\Gamma_0(N)$.
We have also included the character $\chi$ under the modular transformation; where $\chi=1$, the corresponding product is a traditional modular form.
The $[~]$ notation is explained in \eqref{etaprod}. For example, $[1^{24}]$ is simply $\eta(z)^{24} = \Delta(z)$, which is a famous weight 12 modular form.
The two special cases of ``half-integer weight'' are the final two entries.
}}
\label{t:the30}
\end{center}
\end{table}

\subsection{Partition Functions and K3 Surfaces}
Inspired by \eqref{eta24}, we ask whether the shifted reciprocal of all the 30 multiplicative eta-products other than the $\eta(z)^{24}$ have an interesting physical interpretation.
This was partially addressed in the very nice work of \cite{Govindarajan:2009qt}.
The set-up is discussed in detail by the nice review \cite{Sen:2007qy}.

The original context of \cite{Govindarajan:2009qt} was the Chaudhuri-Hockney-Lykken (CHL) maximally supersymmetric heterotic string in less than 10 dimensions \cite{Chaudhuri:1995fk}.
Specifically, \cite{Govindarajan:2009qt} considers asymmetric $\IZ_N$-orbifolds of the $E_8 \times E_8$ heterotic string compactified on the six-torus $T^6 \simeq T^4 \times \tilde{S}^1 \times S^1$.
For our purposes, it is convenient to use string duality to map this to type IIB superstring theory and we shall switch between the two equivalent description liberally.

Considered type IIB compactified on K3 $\times \tilde{S}^1 \times S^1$, which is known to be a 6-dimensional theory with $\cN=4$ supersymmetry.
Now, quotient this theory by a cyclic group $\IZ_t$ action with a generator $g$ acting on the $S^1$ by shifting $1/t$ units along it (i.e., $g = \exp(2 \pi i / t)$ on $S^1$) and simultaneously acting on K3 by an order $N$ involution.
On the heterotic side, within six-torus $T^4 \times \tilde{S}^1 \times S^1$, the $\IZ_t$ acts on the Narain lattice $\Gamma^{20,4}$ associated with the $T^4$ which is a signature $(20,4)$ lattice.
Back from the type II perspective, the lattice can be identified with the cohomology $H^*(\mbox{K3}, \IZ)$. Consequently, the $\IZ_t$ is realized as a so-called {\bf Nikulin involution} which has an Abelian action on K3 that leaves the holomorphic two-form invariant.

Now, consider the configuration consisting of a D5-brane wrapping K3 $\times S^1$, $Q_1$ D1-branes wrapping $S^1$ and Kaluza-Klein monopole with negative magnetic charge associated with $\tilde{S}^1$, $-(k-2)$ units of momentum\footnote{We shift the definition in $k$ to be consistent with our notation.} along $S^1$ and momentum $J$ along $\tilde{S}^1$.
A dyon with  electro-magnetic charge $(q_e, q_m)$, where each is a vector in the lattice $\Gamma^{r,6}$ with $r$ some integer between 1 and 22 as determined by the orbifold action, thus has
\begin{equation}
q_e^2 = 2 (k-2) / t \ , \quad q_m^2 = 2(Q_1-1) \ , \quad q_e \cdot q_m = J \ .
\end{equation}

\subsection{Counting 1/2-BPS States}
In the unorbifolded case, we simply have the heterotic string on $T^6$ and the left-moving sector is bosonic as discussed in \S\ref{s:boson}.
Here, the electric $\frac12$-BPS states carry charge $\frac12 q_e^2$ with $q_e \in \Gamma^{22,6}$.
Level matching gives $n+1 = \frac12 q_e^2$ and the partition function is
\begin{equation}
\frac{16}{\eta(q)^{24}} = \sum\limits_{n=-1}^\infty d_n q^n \ .
\end{equation}
The factor of 16 comes from the Ramond ground state in the right-moving supersymmetric sector and the index $n$ starts at $-1$ due to level matching; therefore, the expression is slightly different from the pure bosonic string given in \eqref{Gbosonic}.

In the orbifolded case, \cite{Govindarajan:2009qt} showed that the above expression generalizes to
\begin{equation}
\frac{16}{\prod\limits_{i=1}^t \eta(n_i z)} = \sum\limits_{n=-1}^\infty d_n q^n \ .
\end{equation}
The factor of 16 is just an overall multiplier.
In all, we have a list of multiplicative partition functions, each associated to a K3 surface with special symplectic automorphism.

Now, Nikulin \cite{nikulin} classified the possible automorphisms of K3 surfaces preserving the holomorphic 2-forms and finite Abelian groups in this list can only be one of the following\footnote{
Incidentally, there are 14 exceptional cases in Arnold's classification of surface singularities and relations between these two lists of 14 were studied in \cite{kobayashi}.
} 14:
\begin{equation}
\IZ_{n = 2, \ldots, 8} \ , \quad \IZ_{m=2,3,4}^2 \ , \quad
\IZ_2 \times \IZ_4 \ , \quad \IZ_2 \times \IZ_6 \ ,\quad 
\IZ_2^{3} \ , \quad \IZ_2^4 \ .
\end{equation}
Consequently, Table 1 of \cite{Govindarajan:2009qt} presents the eta-products which have corresponding Nikulin involutions \footnote{
There is single case of $[11^2,1^2]$ in the last row of their table which is curiously outside the domain of Nikulin involutions, and indeed, as we shall soon see, K3 surfaces.  In the elliptically fibred models, such would have only 4 singular fibres which violates the lower bound of 6.
Nevertheless, the authors have obtained a generating function from the heterotic side.
}.
In particular, they consist of the level $N$ up to 16 eta-products in our Table \ref{t:the30}.
Moreover, as we shall shortly see, they are intimately related to another important set of K3 surfaces.

Before we turn to this next development of our story, we need to emphasize a fact which will be of great utility \cite{nikulin,GS-K3}.
It was shown by Nikulin that the action of the finite Abelian group of symplectic automorphisms is uniquely determined by the integral second cohomology of the K3 surface, which is a lattice of rank 19: $H^2(K3; \IZ) \simeq U^3 \oplus E_8(-1)^2$, where $U$ is a rank 2 hyperbolic lattice and $E_8(-1)$ is a rank 8 negative definite lattice associated to $E_8$.
In other words, the involution does not depend on the specific model of the K3 surface.
Therefore, we can take a convenient algebraic realization in order to perform the necessary computations.
We shall follow \cite{GS-K3-1,GS-K3} and take the K3 surface to be elliptically fibred over $\IP^1$, and in fact with only type-I singular fibres.

\section{K3 Surfaces and Congruence Groups}\label{s:K3}\setall
A similar problem of partitioning 24 arises in the study of K3 surfaces, and through trivalent graphs, congruence subgroups of the modular group.
One cannot resist but to draw analogies to this list and establish a comparative study.
First, let us recall some rudiments.

\subsection{Extremal K3 Surfaces}
Let $X$ be a K3 surface elliptically fibred over a curve $C$, then $C$ is genus 0 and the elliptic $j$-invariant is therefore explicitly a rational map from $C$ to a target $\IP^1$ of degree at most 24.
We will call this the $J$-map.
Explicitly, given the Weierstra\ss\ form of the K3 surface
\begin{equation}
\{ y^2 = 4x^3 -g_2(s) \ x - g_3(s) \} \subset \IC[x,y,s] \ ,
\end{equation}
where $s$ is the affine coordinate of the base curve $C \simeq \IP^1$, the $J$-map is simply, using \eqref{weier}, 
\begin{equation}\label{JmapBelyi}
J = \frac{g_2^3(s)}{\Delta(s)} = \frac{g_2^3(s)}{g_2^3(s) - 27 g_3^2(s)}
\ : \ \IP^1_s \longrightarrow \IP^1 \ .
\end{equation}
Note that we have removed a factor of 1728 from the modular invariant, which we will denote by a lower-case $j$.

In the case of all fibres of $\pi : X \rightarrow C$ being of Kodaira type $I_n$, the K3 surface is called {\bf semi-stable}.
Furthermore, the $J$-map is of degree at most 24, with the {\bf extremal} case of $d=24$ corresponding to the saturation of the Picard number at 20 \cite{shioda-rev}.
How we distribute the singular fibres $I_n$ in the elliptic fibration is then precisely the problem of partitioning $24 = n_1 + \ldots + n_t$.
One of the pioneering papers in this subject is \cite{MP} wherein all possible such distributions, and hence, all extremal, semi-stable elliptic K3 surfaces are classified.
In particular, the number of parts must not be less than 6 by a celebrated result of Shioda-Tate.
The case of $t=6$ partitions is our extremal one.
In all, combining (3.3) to (3.7) of \S3 in \cite{MP} and our Table \ref{t:the30} gives us that all the partitions for $k \ge 3$ exist in both lists (recall that the number of parts is equal to $t = 2k$).

The $J$-maps obey very interesting constraints:
\begin{itemize}
\item 8 preimages of $J(s) = 0$ all having multiplicity (ramification index) 3;
\item 12 preimages of $J(s) = 1$ all having multiplicity (ramification index) 2;
\item $t$ preimages of $J(s) = \infty$, having multiplicity (ramification indices) $[n_1,\ldots,n_t]$;
\item there might be ramification points $x_1, \ldots, x_m$ other than $(0,1,\infty)$ but for $t = 6$, the extremal case, there are no such points.
\end{itemize}
Indeed, Riemann-Hurwitz implies that for ramified covers of $\IP^1 \to \IP^1$, the number of ramification points must exceed the degree of the map by $2 - 2g(\IP^1) = 2$. Here, the former is $8+12+t$ and the latter is 24, whence $t=6$, the extremal case, is the only one for which there are no other ramification points other than $(0,1,\infty)$.

Now, maps to $\IP^1$ ramified only at $(0,1,\infty)$ hold a crucial place in modern number theory and are call {\bf Belyi maps}.
Thus, for our 6-tuple partitions, the J-maps are Belyi.
These can be represented graphically as Grothendieck's {\bf dessins d'enfants}. 
To draw such a dessin is simple: given the ramification data
$\left\{ (\vec{r}_0)_{i=1,\ldots,W}, (\vec{r}_1)_{j=1,\ldots,B}, (\vec{r}_\infty \right)_{k=1,\ldots,I}\}$ specifying the ramification indices at the various pre-images of 0, 1 and infinity, one marks one white node for the $i$-th pre-image of 0, with $(r_0)_i$ edges emanating therefrom; similarly, one marks one black node for the $j$-th pre-image of 1, with $(r_1)_j$ edges. Thus we have a bipartite graph embedded on a Riemann sphere, with W white nodes and B black nodes. Now we connect the nodes with the edges, joining only black with white, such that we have I faces, each being a polygon with $(2r_\infty)_k$ sides.

In our present case, the ramification data is $\left\{3^8, 2^{12}, [n_1,\ldots,n_t] \right\}$. Note that such dessins are called {\bf clean} because all pre-images of 1 have valency 2.
The dessins for all the extremal 6-tuple cases are studied in \cite{BM,He:2012jn} and we refer the reader to Appendix A of \cite{He:2012jn}.

\subsection{Modular Subgroups and Coset Graphs}
In \cite{mckaysebbar,sebbar}, a particular family of subgroups $G$ of the modular group $\Gamma = PSL(2;\IZ)$  has been identified. These are the so-called {\bf torsion-free} and {\bf genus zero} congruence subgroups.
By torsion-free we mean that the subgroup contains no element, other than the identity, which is of finite order.
By genus zero we mean that when we quotient the upper half plane $\cH$ (compactified to $\cH^*$ by adjoining so-called ${\it cusps}$, which are points on $\IQ \cup \infty$) by the subgroup $G$, the resulting Riemann surface is genus 0.
Indeed, $\cH^*$ quotiented by the full modular group $\Gamma$ is well-known to be a Riemann sphere.

Now, the stabilizer of the cusp is a finite index subgroup of $G$, which is a finite index $n$ subgroup of the stabilizer of the cusp in the full $\Gamma$; we call $n$ the cusp width associated with the cusp for $G$.
It is also the smallest positive integer such that the modular conjugate of the action $z \mapsto z + n$ leaves the cusp invariant.
The sum over the cusp widths turns out to be the index of the subgroup $G$ itself in $\Gamma$.

The complete classification of the torsion-free, genus zero, subgroups of $\Gamma$ was carried out in \cite{classSebbar} and they are very rare indeed: 
they are only 33, all of index 6, 12, 24, 36, 48 or 60.
In particular, there are 9 of index 24, and the relation to gauge theories was discussed in \cite{He:2012kw}.
Given the aforementioned cusp widths, these 9 groups will correspond to 6-tuple partitions of 24, as given in Table \ref{t:Nineg=0}.
\begin{table}[h!!!]
\begin{center}
\[
\begin{array}{cc}
\begin{array}{|l|l|}\hline
\hline
\mbox{Group} & \mbox{Cusp Widths} \\ \hline \hline
$Ia$ : \Gamma(4) & [4^6] \\ \hline
$Ib$ : \Gamma(8; 4,1,2) & [2^2, 4^3, 8] \\ \hline
$IIa$ : \Gamma_0(3) \cap \Gamma(2) & [2^3, 6^3] \\ \hline
$IIb$ : \Gamma_0(12) & [1^2, 3^2, 4, 12] \\ \hline
$IIIa$ :  \Gamma_1(8) & [1^2, 2, 4, 8^2] \\ \hline
$IIIb$ : \Gamma_0(8) \cap \Gamma(2) & [2^4, 8^2] \\ \hline
$IIIc$ : \Gamma_0(16) & [1^4, 4, 16] \\ \hline
$IIId$ : \Gamma(16;16,2,2) & [1^2, 2^3, 16] \\ \hline
$IV$ : \Gamma_1(7) & [1^3, 7^3]
\\ \hline
\end{array}
&
\begin{array}{l}
\mbox{ where } \\
~\\
\Gamma(m) := \{A \in SL(2;\IZ)  \, \left| \, A \equiv \pm I \bmod m \right.\} 
/ \{\pm I \}\\
\Gamma_1(m) := \left\{A \in SL(2;\IZ)  \, \left| \, A \equiv \pm \left( \begin{matrix}
1 & b \\ 0 & 1
\end{matrix} \right) \bmod m \right. \right\} / \{\pm I \} \ ;\\
\Gamma_0(m) := \left\{ \left. \left( \begin{matrix}
a & b \\ c & d
\end{matrix} \right) \in \Gamma  \, \right| \, c \equiv 0 \bmod m \right\} / \{\pm I \}\\
\Gamma(m; \frac{m}{d}, \epsilon, \chi) :=
\left\{ \left.
\pm \left(\begin{matrix}
1 + \frac{m}{\epsilon \chi} \alpha & d \; \beta\\
\frac{m}{\chi} \gamma & 1 + \frac{m}{\epsilon \chi} \delta
\end{matrix}\right) \, \right| \,
\gamma \equiv \alpha \bmod \chi 
\right\} \ .
\end{array}
\end{array}
\]
\caption{{\sf
The 9 torsion free, genus zero, congruence subgroups of the modular group.
}}
\label{t:Nineg=0}
\end{center}
\end{table}

Now, each of these is an index 24 subgroup of the modular group, and we can draw the Schreier coset graph for each.
First, recall that the Cayley graph of $PSL(2;\IZ)$ is an infinite free trivalent tree, but with each node replaced by an oriented triangle.
This is because $\Gamma := PSL(2;\IZ) \simeq \langle S, T  \, \left| \, S^2 = (ST)^3 = I \rangle \right.$; calling $x$ the element of order 2 and $y$ the element of order 3, we see that $\Gamma$ is the free product of the cyclic groups $C_2 = \gen{x | x^2 = I}$ and $C_3 = \gen{y | y^3 = I}$.
That is, $\Gamma \simeq C_2 \star C_3$.
Thus $x$ will serve as an undirected edge whilst $y$ will give rise to an oriented triangle, namely, a directed triangular closed circuit.

For a subgroup $G \in PSL(2;\IZ)$ of index $\mu$, we can decompose the modular group into the (right) cosets $G g_i$ of $G$ as 
$PSL(2;\IZ) \simeq \bigcup\limits_{i=1}^\mu G g_i$,
so that our generators $x$ and $y$ act by permuting the nodes, which now correspond to cosets.
The result is a coset graph with $\mu$ nodes and a folded version of the Cayley graph of the full modular group.
This is the {\bf Schreier coset graph} (sometimes called Schreier-Cayley coset graph) and
it remains, in particular, to be trivalent, with bi-directional edges for $x$ and oriented 3-cycles for $y$.
In fact, the converse is true: {\it any} finite cubic graph is a realization of a Schreier coset graph of a subgroup of the modular group.

To complete the story, we can canonically associate a K3 surface to each of these genus zero subgroups.
First, we extend the action of $G \subset \Gamma$  on $\cH$ to an action
\begin{equation}
\cH \times \IC \ni (z,w)
\longrightarrow \left( \gamma z, \frac{w + m z + n}{c z + d}
\right) \ ,
\end{equation}
for $\gamma = \mat{a & b \\ c & d} \in G$  and $(m,n) \in \IZ^2$.
Thus the quotient of $\cH \times \IC$ by the above automorphism defines a surface equipped with a morphism to the modular curve arising from the quotient of $\cH$ by $z \to \gamma z$.
The fibre over the image of this morphism to the modular curve is generically an elliptic curve corresponding to the lattice $\IZ \oplus \IZ \tau_T$ with complex structure parametre $\tau_T$.
What we have therefore is a complex surface which is an elliptic fibration over the modular curve, called the {\bf Shioda elliptic modular surface} \cite{shioda-rev} associated to $G$.
The base, because our modular curves are genus zero, will be the Riemann sphere $\IP^1_C$. 

For our index 24 subgroups $G$, the modular surface is a semi-stable, extremal, elliptic K3 surface, the 6 cusp widths are precisely the 6 $I_n$ fibres.
Moreover, the Schreier coset graph $G$ is, when replacing each oriented triangle with a black node and inserting in each edge a white node, the dessin d'enfant for the $J$-map of the corresponding K3 surface \cite{He:2012kw}.

\subsection{Summary}\label{s:summary}
In summary, we present the objects which lie in the intersection of all the above partitioning problems of 24 in Table \ref{t:sum}.
Starting from the left, the first column is the cycle shape of the eta-product $[n_1, \ldots, n_t]$ as defined in \eqref{etaprod}.
Next, in column 2, we have the weight $k$, level $N$ and the character $\chi$ under which the eta-product transforms modularly as in \eqref{modulartrans} and Table \ref{t:the30}.
The eta-product is the partition function of certain quotients of the type IIB string theory compactified on K3$\times T^2$ with special K3 surfaces admitting Nikulin involution as given in column 4.
The cycle shape, being a partition of 24, also uniquely determines an extremal K3 surface which is semi-stable with type-$I_n$ fibers exactly being $\{I_{n_1}, \ \ldots, I_{n_t} \}$.
These K3 surfaces are elliptically fibred over $\IP^1$, with $j$-invariants being rational functions in the homogeneous coordinate $s$ of the base $\IP^1$, given in column 5.
They can be considered as ramified maps from $\IP^1$ to $\IP^1$, which turn out to be Belyi, and hence describe clean dessins d'enfants, as drawn in column 6.
The dessins are precisely Schreier coset graphs (column 6) associated with congruence subgroups (column 3) of the modular group $PSL(2;\IZ)$.

\begin{table}[t!!!]
\[
\hspace{-0.7in}
\begin{array}{|c|c|c|c|c|c|} \hline
\mbox{Eta Product} & (k,N,\chi) & 
\mbox{\begin{tabular}{l}Modular \\Subgroup\end{tabular}} & 
\mbox{\begin{tabular}{l}Nikulin \\Involution\end{tabular}} & 
J\mbox{-Map} &
\mbox{\begin{tabular}{l}Dessin \& \\Schreier\end{tabular}}
\\ \hline \hline
[7^3,1^3] & (3,7,\jacobi{-7}{d}) & \Gamma_1(7) & \IZ_7 
& 
\frac{\left(s^8-12 s^7+42 s^6-56 s^5+35 s^4-14 s^2+4 s+1\right)^3}{(s-1)^7 s^7
   \left(s^3-8 s^2+5 s+1\right)}
& 
\begin{array}{c} \includegraphics[trim=0mm 0mm 0mm 0mm, clip, width=1.0in]{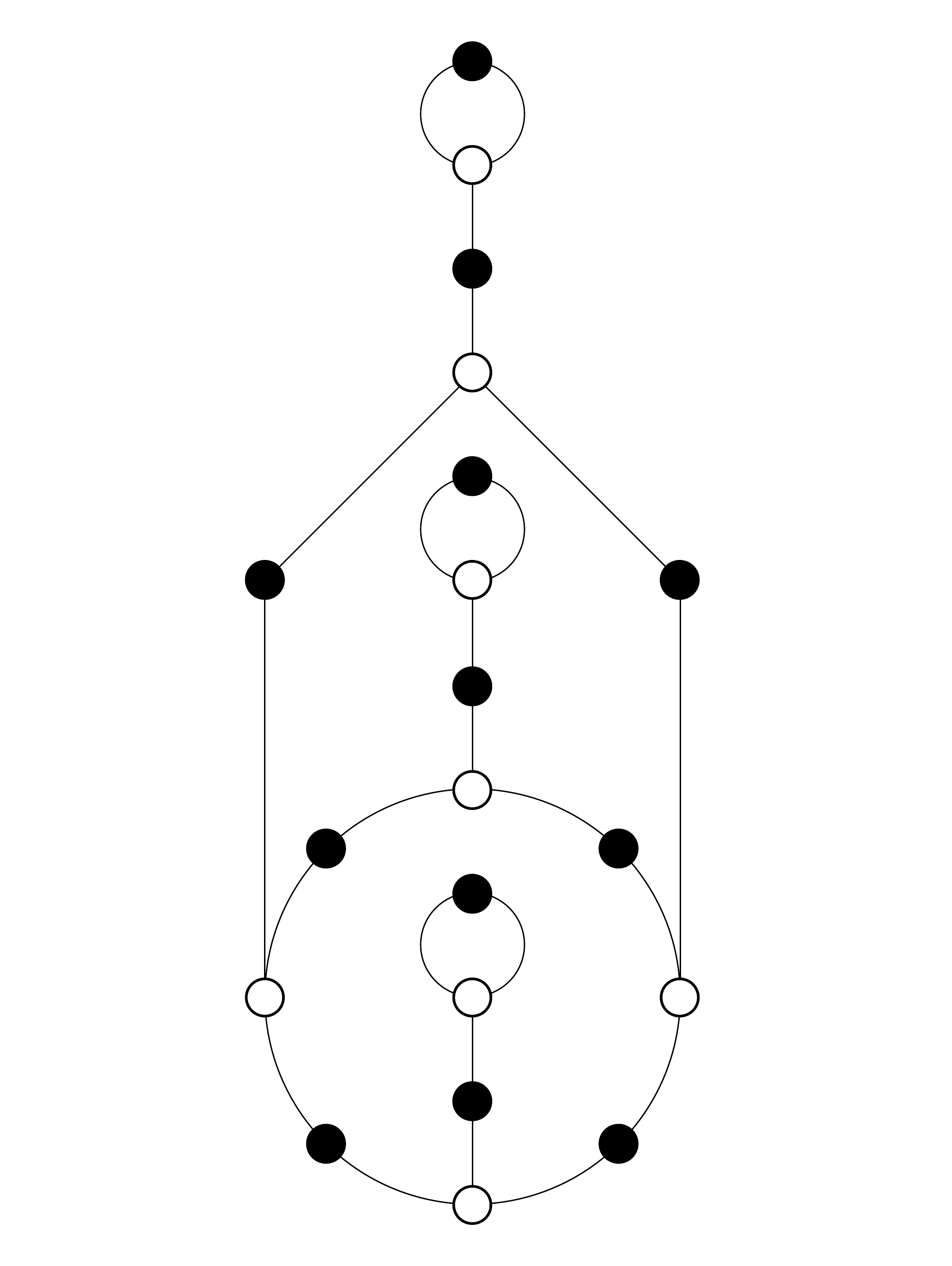}  \end{array}
\\ \hline
[8^2,4,2,1^2] & (3,8,\jacobi{-2}{d}) & \Gamma_1(8) & \IZ_8
&
-\frac{16 \left(s^8-28 s^6-10 s^4+4 s^2+1\right)^3}{s^4 \left(s^2+1\right)^8 
\left(2s^2+1\right)}
&
\begin{array}{c} \includegraphics[trim=0mm 0mm 0mm 0mm, clip, width=1.0in]{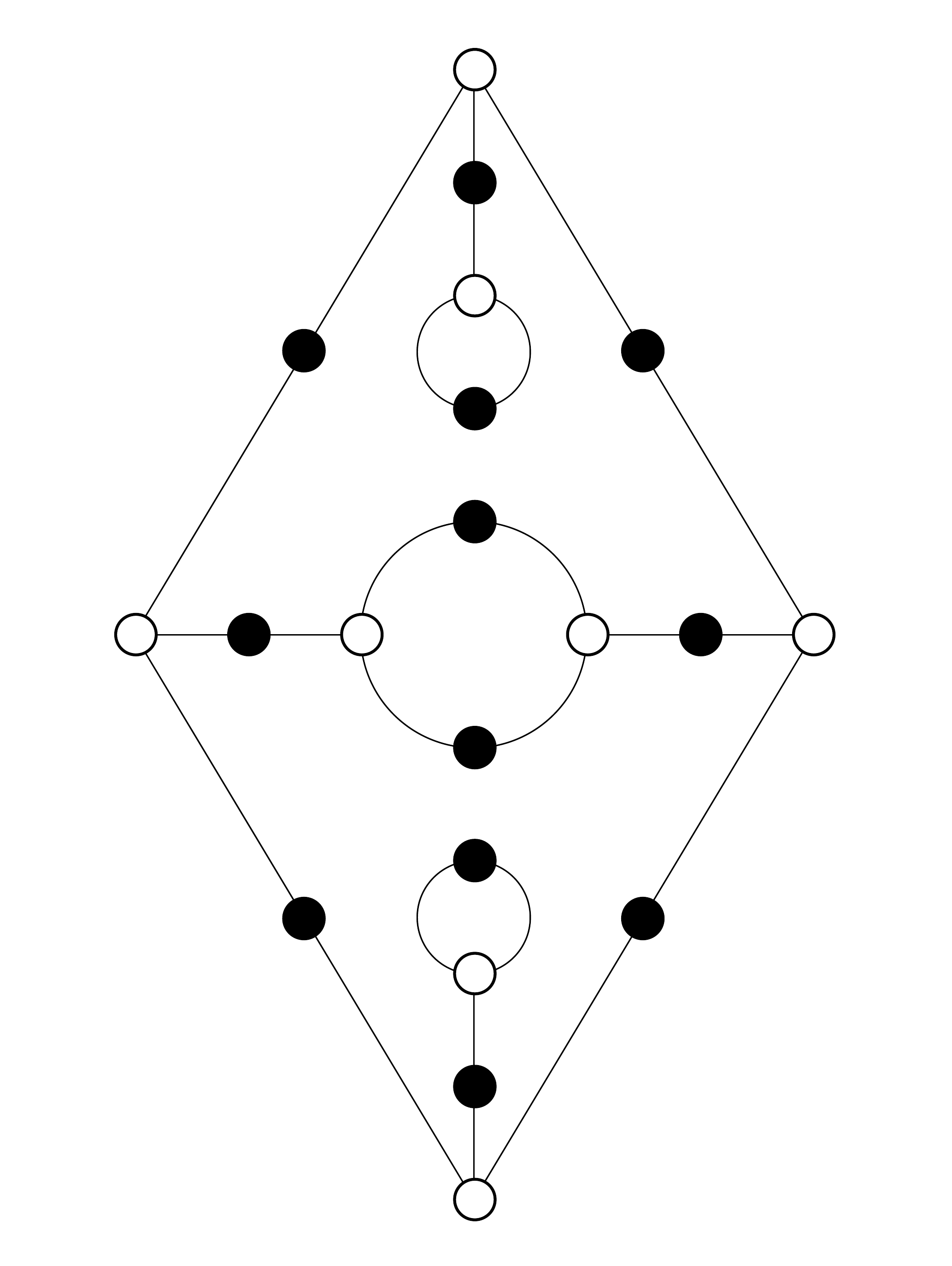}  \end{array}
\\ \hline
[6^3,2^3] & (3,12,\jacobi{-3}{d} ) & \Gamma_0(3) \cap \Gamma(2)  &
\IZ_2 \times \IZ_6
&
\frac{\left(3 s^2+8\right)^3 \left(3 s^6+600 s^4-960 s^2+512\right)^3}{8 s^6 
\left(8-9s^2\right)^2 \left(s^2-8\right)^6}
&
\begin{array}{c} \includegraphics[trim=0mm 0mm 0mm 0mm, clip, width=1.0in]{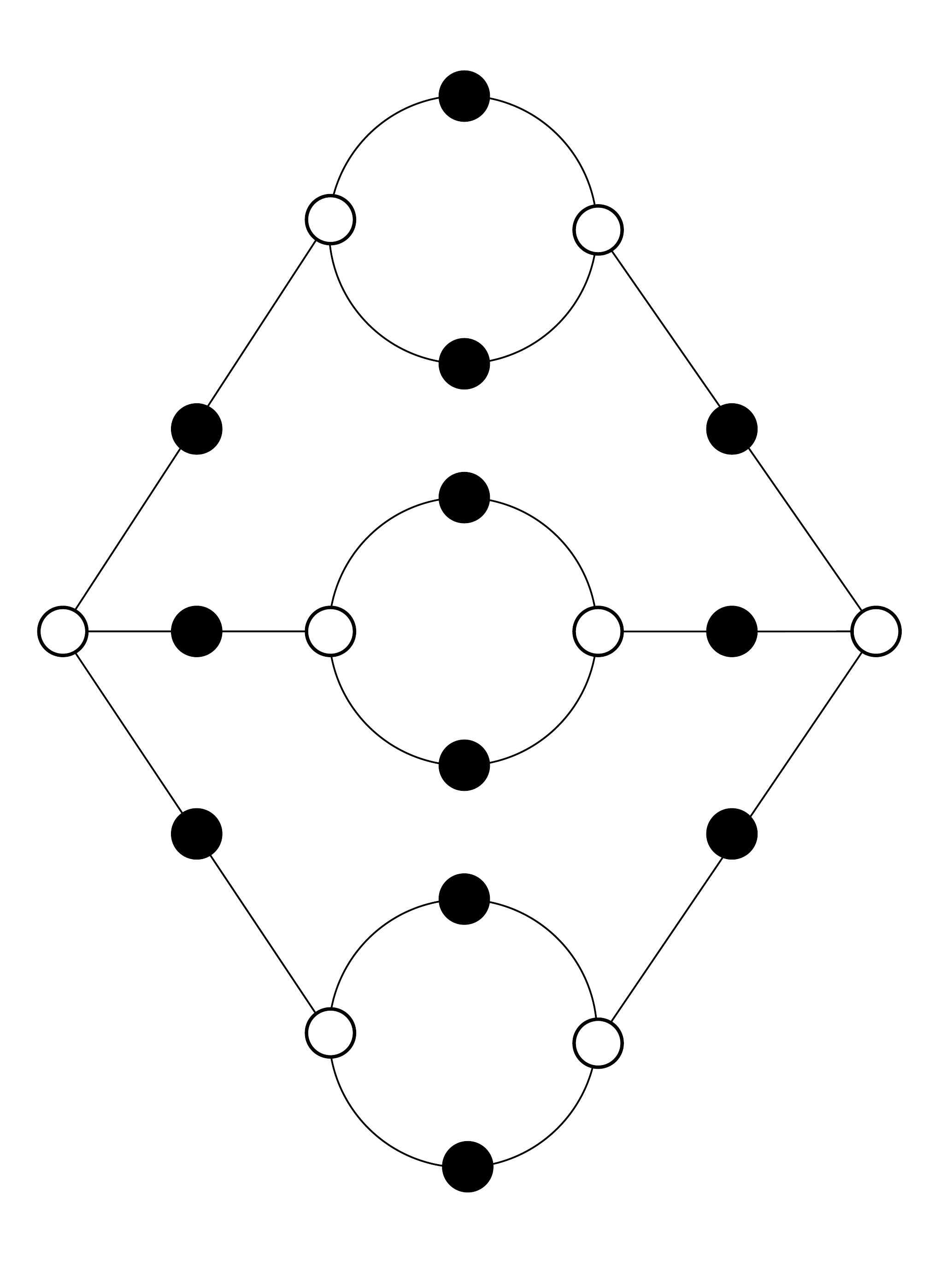}  \end{array}
\\ \hline
[4^6] & (3,16,\jacobi{-1}{d}) & \Gamma(4) & \IZ_4^2
&
\frac{16(1+14s^4+s^8)^3}{s^4(s^4-1)^4}
&
\begin{array}{c} \includegraphics[trim=0mm 0mm 0mm 0mm, clip, width=1.0in]{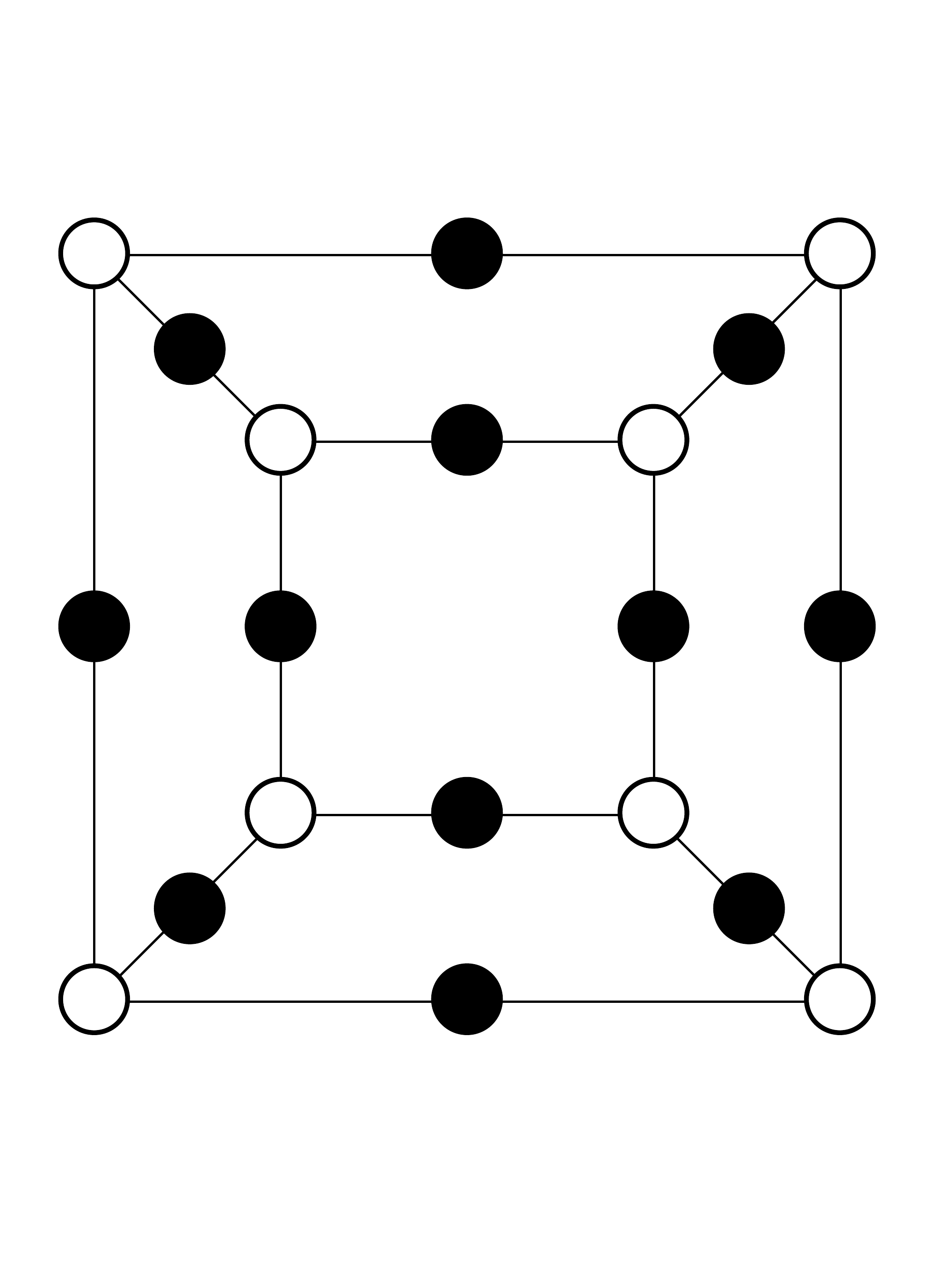}  \end{array}
\\ \hline
\end{array}
\]
\caption{{\sf
The four eta-products relevant to extremal K3 surfaces in two contexts: modular elliptic K3 surfaces/dessins/congruence subgroups and type IIB compactification on K3$\times T^2$/Nikulin involutions/partition functions; q.v., text in \S\ref{s:summary} for an explanation of the various columns.
}}
\label{t:sum}
\end{table}

To complete the cycle of correspondences, we know the explicit equations of the K3 surfaces from both sides:
(1) From the point of view of the modular surfaces, the Weierstra\ss\ form of the elliptic fibration has been computed in \cite{TopYui,He:2012kw} which yields the correct J-maps in Table \ref{t:sum};
(2) From the perspective of the partition function on K3$\times T^2$, the K3 surfaces admitting Nikulin involutions also have explicit models in Tate, Weierstra\ss\ or Legendre form, which are computed in \cite{GS-K3}.
For the ones of our interest as given in Table \ref{t:sum}, we summarize the equations in Table \ref{t:compare}, wherein $s$ is the base projective coordinate of the $\IP^1$ over which the K3 surface is an elliptic fibration.
In the second column of the equations for the Nikulin K3, $p$ and $q$ are some linear functions $a s +b$ (with $a,b \in \IC$) in $s$ and not the same for each of the cases.

\begin{table}[t!!!]
\[
\hspace{-0.7in}
\begin{array}{|c|c||c|c|}\hline
\mbox{Nikulin Inv} & \mbox{Equation} & \mbox{Congruence Group} & \mbox{Equation} \\ \hline \hline
\IZ_7 & 
\begin{array}{l} y^2 + (1+s-s^2) x y + (s^2-s^3)y \\
\qquad \qquad = x^3 + (s^2-s^3) x^2 
\end{array}
& \Gamma_1(7) & \mbox{same} 
\\ \hline

\IZ_8 & 
y^2 = x^3 + (\frac{(p-q)^4}{4} - 2p^2 q^2)x^2 + p^4 q^4 x 
& \Gamma_1(8) & 
(x+y)(xy-1) + \frac{4i s^2}{s^2+1}xy = 0
\\ \hline

\IZ_2 \times \IZ_6 & 
\begin{array}{l} 
y^2 = x\left(x- (3p-q)(p+q)^3\right) \times\\
\qquad \qquad \left(x -  (3p+q)(p-q)^3 \right)
\end{array}
& \Gamma_0(3) \cap \Gamma(2) & 
\begin{array}{l}
(x+y)(x+1)(y+1) +  \\
\qquad \qquad \frac{8s^2}{8-s^2} xy = 0
\end{array}
\\ \hline

\IZ_4^2 & 
y^2 = x\left(x-p^2q^2\right)\left(x-\frac{(p^2+q^2)^2}{4}\right)
& \Gamma(4) & 
\begin{array}{l}
x(x^2+2y+1) + \\
\qquad \frac{s^2-1}{s^2+1}(x^2-y^2) = 0
\end{array}
\\ \hline
\end{array}
\]
\caption{{\sf
The explicit equations of the K3 surfaces: from Nikulin's list and from the modular extremal list.
In all cases, $s$ is the base projective coordinate of the $\IP^1$ over which the K3 surface is an elliptic fibration.
The coefficients $p$ and $q$ are some (not the same for each case) linear functions of $s$. 
}}
\label{t:compare}
\end{table}

We see that in the first case of the modular K3 surface associated to $\Gamma_1(7)$ the one with the $\IZ_7$ involution has the identical equation.
This is a semi-stable extremal K3 with 6 type $I_n$ fibres:  three each of $I_7$ and $I_1$, which is precisely the cycle shape of the corresponding eta product.
Incidentally, this is an elliptic curve known for some time \cite{kubert}.

For the remaining three cases, we have degrees of freedom from the linear functions $p$ and $q$, which can be fixed by appropriate transformations to the forms from the modular side.
The easiest strategy is to simply compute the J-invariant for the equations from the Nikulin side and match to the J-invariant as given in column 5 of Table \ref{t:sum}.
Now, it is a standard fact that for an elliptic curve in Tate form
\begin{equation}
y^2 + a_1 x y + a_3 = x^3 + a_2 x^2 + a_4 x + a_6 \ ,
\end{equation}
the J-invariant (without the 1728 prefactor) is given by
\begin{align}
\nn
J &= \frac{c_4^3}{-b_2^2b_8+9b_2b_4b_6-8b_4^3-27b_6^2} \ , \mbox{ with }
\\
\label{JTate}
&
\begin{array}{l}
b_2 = a_1^2+4a_2,\quad b_4=a_1a_3+2a_4,\\
b_6=a_3^2+4a_6,\quad b_8=a_1^2a_6-a_1a_3a_4+a_2a_3^2+4a_2a_6-a_4^2,\\
c_4 = b_2^2-24b_4,\quad c_6 = -b_2^3+36b_2b_4-216b_6 \ .
\end{array}
\end{align}

For $\IZ_4^2$, the J-invariant is, using \eqref{JTate} and Table \ref{t:compare}, $\frac{16 \left(p^8+14 p^4 q^4+q^8\right)^3}{p^4 q^4\left(p^4-q^4\right)^4}$.
Therefore, comparing with the J-map from Table \ref{t:sum}, this sets $p=s$ and $q=1$, which are indeed linear in $s$.
Consequently, the specific Nikulin K3 surface becomes $y^2=x(x-s^2)(x-\frac{(s^2+1)^2}{4})$.

For $\IZ_8$, if we perform a simple change of base variables $s \rightarrow 1/s$ for the congruence group/modular elliptic curve, giving us a $J$-invariant
$-\frac{16 \left(s^8-28 s^6-10 s^4+4 s^2+1\right)^3}{s^4
   \left(s^2+1\right)^8 \left(2 s^2+1\right)}$, then setting $p = s+i$ and $q = s-i$ gives precisely this expression using \eqref{JTate}; again $p$ and $q$ are linear functions, as required.
Hence, here the K3 surface is described by $y^2=x^3+(4 - 2(s^2+1)^2)x^2+(s^2+1)^4x$.

Finally, for $\IZ_2\times\IZ_6$, we find from Table \ref{t:compare} that $J = \frac{\left(9 p^8+228 p^6 q^2+30 p^4 q^4-12 p^2 q^6+q^8\right)^3}{\left(p^3-p q^2\right)^6 \left(q^3-9 p^2 q\right)^2}$.
Setting $p = s$ and $q = \sqrt{8}$ immediately gives the J-map of the corresponding modular K3 surface in Table \ref{t:sum}.
Therefore, the K3 surface is given by $y^2=x^3+\left(-6 s^4-96 s^2+128\right) x^2+\left(9 s^8-224 s^6+1920 s^4-6144s^2+4096\right) x$.

Therefore, we conclude that we are indeed talking about the same K3 surfaces, both from the modular elliptic/Cayley graph side and from the Nikulin involution/BPS state counting side. In the latter, we are fixed at particular points in the space of complex structure, since in the former, there is complete rigidity because of the algebraic nature of dessins d'enfants.

\subsection{Beyond Extremality}
We have discussed the case of 6 type-I fibres extensively so far, which, as mentioned above, correspond to extremal K3 surfaces; of course, both the cycle shape of the eta-products and of the Nikulin involutions can exceed the lower bound of 6.
Examining Table \ref{t:the30}, there are 9 cases of $k > 3$.
The cycle shapes of these, re-reassuringly, also all - except the maximal case of $k=12$ - appear in the list of \cite{GS-K3} (for the cases of $\IZ_3$ and $\IZ_5$, the full equations were given in a preceding work \cite{GS-K3-1}), as sequences of $I_n$ fibres.
The explicit Weierstra\ss\ equations are presented in Table \ref{t:Nik}.
In it, we have adhered to the notation that $p_i$ and $q_i$ are some degree $i$ polynomial in the base coordinate $s$.

\begin{table}[t!!!]
\begin{center}
\[
\begin{array}{|c|c|c|c|} \hline
\mbox{Eta Product} & (k,N,\chi) & 
\mbox{\begin{tabular}{l}Nikulin \\Involution\end{tabular}} & 
\mbox{Equation}
\\ \hline \hline
 
[2^8, 1^8] & (8,2,1) & \IZ_2 &  y^2 = x(x^2 + p_4 x + q_8)
 \\ \hline

[3^6,1^6] & (6,3,1) & \IZ_3 & y^2 = x^3 + \frac13 x (2p_2q_6 + p_2^4) + 
  \frac{1}{27} (q_6^2 - p_2^6)
 \\ \hline

[2^{12}] & (6,4,1) & \IZ_2^2 & y^2 = x(x-p_4)(x-q_4) 
 \\ \hline

[4^4,2^2,1^4] & (5,4,\jacobi{-1}{d}) & \IZ_4 & 
  y^2 = x(x^2+ (p_2 - 2q_4) x + q_4^2)
 \\ \hline

[6^2,3^2,2^2,1^2] & (4,6,1) & \IZ_6 & 
  y^2 = x(x^2 + (-3p_2^2 + q_2^2) x + p_2^3(3p_2 + 2q_2))
 \\ \hline

[5^4,1^4] & (4,5,1) & \IZ_5 & 
\begin{array}{l}
  y^2 = x^3 + \frac13x\left(-q_2^4 + p_2^2q_2^2 - p_2^4 - 
        3p_2q_2^3 + 3p_2^3q_2\right) + \\
  \quad + \frac{1}{108}(p_2^2+q_2^2)(19q_2^4-34p_2^2q_2^2+19p_2^4+18p_2q_2^3-18p_2^3q_2)
\end{array}
 \\ \hline

[4^4,2^4] & (4,8,1) & \IZ_2 \times \IZ_4 &
  y^2 = x(x-p_2^2)(x-q_2^2) 
 \\ \hline

[3^8] & (4,9,1) & \IZ_3^2 &
\begin{array}{l}
  y^2 = x^3 + 12 x \left( (s^2+1)(p_0 s^2 + q_0)^3 + (s^2+1)^4 \right) + \\
\quad + 2 \left( (p_0 s^2 + q_0)^6 - 20(p_0 s^2 + q_0)^3(s^2+1)^3 - 8(s^2+1)^6
      \right)
\end{array}
 \\
\hline
\end{array}
\]
\caption{{\sf
The K3 surfaces which admit Nikulin involutions which correspond to non-extremal cases (the number $k/2$ of partitions of 24 is not equal to 6).
In the explicit Weierstra\ss\ equation, $p_i$ and $q_i$ are some degree $i$ polynomials in the base coordinate $s$.
For reference, we record the corresponding eta-product, as well as its level $N$ and character $\chi$ as a modular form.
}}
\label{t:Nik}
\end{center}
\end{table}

Once again, we can find algebraic points in the moduli space of these K3 surfaces which make the J-maps Belyi.
As an example, let us look at $[3^8]$.
Using \eqref{JTate}, we readily see that the j-invariant (with the 1/1728 factor) is $j_{[3^8]}(s) =$
\begin{equation}
{\scriptsize
\frac{64 \left(\left(s^2+1\right) \left({p_0}
   s^2+{q_0}\right)^3+\left(s^2+1\right)^4\right)^3}{64
   \left(\left(s^2+1\right) \left({p_0}
   s^2+{q_0}\right)^3+\left(s^2+1\right)^4\right)^3+\left(-20
   \left(s^2+1\right)^3 \left({p_0}
   s^2+{q_0}\right)^3+\left({p_0} s^2+{q_0}\right)^6-8
   \left(s^2+1\right)^6\right)^2} \ .
}
\end{equation}
Seeing that the discriminant of the numerator of $j-1$ vanishes is reassuring: it is indeed Belyi.
For example, setting $p_0=1, q_0=0$ gives us a Belyi map with 8 pre-images of 0 with ramification 3 and 12 pre-images of 1 with ramification 12.
We could, of course, perform a similar exercise for all the remaining non-extremal cases, but for now, we seem to have exhausted K3 surfaces with the weight $k \ge 3$ (and hence cycle length $\ge 6$) cases, we now turn to the $k=2$ eta-products.

\subsection{Elliptic Curves}\label{s:modular}\setall
Now, in the list in Table \ref{t:the30}, there are ones of weight $k=2$ which are modular forms at various levels.
According to the celebrated theorem of Taniyama-Shimura-Wiles, these should be associated to some elliptic curve in the sense that the Hasse-Weil L-function should be the Mellin transform of these eta-products; moreover the conductor of the elliptic curve should be the level.
Such a situation - and in fact, more generally allowing quotients of eta functions as well - was considered in \cite{ono}.
The reader is also referred to \cite{He:2010mh,He:2011ge} for L-functions in the context of gauge theories, especially in light of the Plethystic programme.

Specifically, defining the standard Tate form of an elliptic curve as
\begin{equation}
y^2 + a_1 xy + a_3 y = x^3 + a_2 x^2 + a_4 + a_6 \ ,
\end{equation}
we have the correspondence (we reproduce their table here, and also include the $j$-invariant, with the 1/1728 normalization):
\begin{equation}
\begin{array}{|c|c|c|c|} \hline
N & \mbox{eta-product} & (a_1,a_2,a_3,a_4,a_6) & j \\ \hline \hline
15 & [15,5,3,1] & (1,1,1,-10,-10) & 13^3 \cdot 37^3 / 2^6\cdot 3^7 \cdot 5^4 \\
14 & [14,7,2,1] & (1,0,1,4,-6) & 5^3 \cdot 43^3 / 2^{12} \cdot 3^3 \cdot 7^3 \\
24 & [12,6,4,2] & (0,-1,0,-4,4) & 13^3 / 2^2 \cdot 3^5  \\
11 & [11^2,1^2] & (0,-1,1,-10,-20) & -2^6 \cdot 31^3 / 3^3 \cdot 11^5 \\
20 & [10^2,2^2] & (0,1,0,4,4) & 11^3 / 2^2 \cdot 3^3 \cdot 5^2 \\
27 & [9^2,3^2] &  (0,0,1,0,-7) & 0 \\
32 & [8^2,4^2] & (0,0,0,4,0) & 1 \\
36 & [6^4] & (0,0,0,0,1) & 0\\
\hline
\end{array}
\end{equation}
It is interesting to see that $N=27$ and $N=36$ are isogenous and correspond to the simple elliptic curve $y^2 = x^3 + 1$.
In general, our multiplicative product \cite{FM} affords a Mellin transform which is an Euler product over L-functions as
\begin{align}
\nn
M\left([ n_1, n_2, \ldots, n_t ] \right) &= \prod\limits_{p\mbox{ {\scriptsize prime}}} 
\left(1 - a_p p^{-s} + b_p p^{-2s} \right)^{-1} \\
b_p &:= \jacobi{-N}{p}^k p^{k-1}
\ ,
\end{align}
where the conductor $N$ is the product of the largest and smallest entries in the cycle shape (i.e., $n_1 n_t$ if $[ n_1, n_2, \ldots, n_t ] $ is ordered) and $k$ is, as always, the half-cycle-length, which is also the weight of the modular form.

Indeed, the q-expansions of these eta-products as modular forms should have multiplicative coefficients, much in the spirit of $\eta(q)^{24}$ discussed at the very beginning of our exposition, which is not in the present list because it is of weight ``$\frac12$''.
Nevertheless, to give an idea to the explicit q-expansions, we see that
\begin{align}
\nn
[15,5,3,1] = &
q-q^2-q^3-q^4+q^5+q^6+3 q^8+q^9-q^{10}-4 q^{11}+q^{12}-2
   q^{13}-q^{15}-q^{16}+2 q^{17}-\\ \nn
& -q^{18}+4 q^{19}-q^{20}+4 q^{22}-3q^{24}+q^{25}+2 q^{26}+\cO\left(q^{27}\right)\\
\nn
[14,7,2,1] =&
q-q^2-2 q^3+q^4+2 q^6+q^7-q^8+q^9-2 q^{12}-4 q^{13}-q^{14}+q^{16}+6
   q^{17}-q^{18}+\\ \nn
&+2 q^{19}-2 q^{21}+2 q^{24}-5 q^{25}+4 q^{26}+4
   q^{27}+q^{28}-6 q^{29}+\cO\left(q^{30}\right)
\\ \nn
[12,6,4,2] =&
q-q^3-2 q^5+q^9+4 q^{11}-2 q^{13}+2 q^{15}+2 q^{17}-4 q^{19}-8
   q^{23}-q^{25}-q^{27}+6 q^{29}+\\ \nn
& + 8 q^{31}-4 q^{33}+6 q^{37}+2 q^{39}-6
   q^{41}+4 q^{43}-2 q^{45}-7 q^{49}-2 q^{51}-2 q^{53}-8 q^{55}+\cO\left(q^{57}\right)
\\ \nn
[11^2,1^2] =&
q-2 q^2-q^3+2 q^4+q^5+2 q^6-2 q^7-2 q^9-2 q^{10}+q^{11}-2 q^{12}+4
   q^{13}+4 q^{14}-q^{15}-
\\ \nn
& -4 q^{16}-2 q^{17}+4 q^{18}+2 q^{20}+2 q^{21}-2
   q^{22}-q^{23}-4 q^{25}-8 q^{26}+5 q^{27}-4 q^{28}+\cO\left(q^{30}\right)
\\ \nn
[10^2,2^2] =&
q-2 q^3-q^5+2 q^7+q^9+2 q^{13}+2 q^{15}-6 q^{17}-4 q^{19}-4 q^{21}+6
   q^{23}+q^{25}+4 q^{27}+6 q^{29}- \\ \nn
& 4 q^{31}-2 q^{35}+2 q^{37}-4 q^{39}+6
   q^{41}-10 q^{43}-q^{45}-6 q^{47}-3 q^{49}+12 q^{51}-6 q^{53}
   +\cO\left(q^{54}\right)
\\ \nn
[9^2,3^2] =&
q-2 q^4-q^7+5 q^{13}+4 q^{16}-7 q^{19}-5 q^{25}+2 q^{28}-4 q^{31}+11
   q^{37}+8 q^{43}-\\ \nn
& -6 q^{49}-10 q^{52}-q^{61}-8 q^{64}+5 q^{67}-7
   q^{73}+14 q^{76}+17 q^{79}-5 q^{91}-19 q^{97}+O\left(q^{100}\right)
\\ \nn
[8^2,4^2] = &
q-2 q^5-3 q^9+6 q^{13}+2 q^{17}-q^{25}-10 q^{29}-2 q^{37}+10 q^{41}+6
   q^{45}-7 q^{49}+14 q^{53}-
\\ \nn
& -10 q^{61}-12 q^{65}-6 q^{73}+9 q^{81}-4
   q^{85}+10 q^{89}+18 q^{97}+\cO\left(q^{98}\right)
\\ \nn
[6]^4 =& q-4 q^7+2 q^{13}+8 q^{19}-5 q^{25}-4 q^{31}-10 q^{37}+8 q^{43}+9
   q^{49}+14 q^{61}-\\ \nn
& - 16 q^{67}-10 q^{73}-4 q^{79}-8 q^{91}+14
   q^{97}+\cO\left(q^{101}\right)
\end{align}
Indeed, the multiplicativity of the initial coefficients is evident.

The natural course of action, of course, is to take the Dirichlet transform $L(s) = \sum\limits_{n=1}^\infty a_n n^{-s}$ of these multiplicative coefficients $a_n$, which, by the Modularity Theorem, should be the L-function of the corresponding elliptic curve.
For example, take the simple case of $[6]^4$, the elliptic curve is $y^2 = x^3 + 1$, whose local zeta-function can be computed - by Magma \cite{magma} for instance - and taking the product over the primes (both of good and bad reduction) indeed gives the coefficients in the last row above. 
The explicit forms of the local zeta-functions, depending on the prime $p$, can be readily given as rational functions by Weil-Deligne in the standard way:
here the conductor is 36, thus the global zeta-function is equal to
\begin{equation}
Z(s) = \frac{\zeta(s)\zeta(s-1)}{L(s)} = \prod_{p \nmid 36} \frac{1 - 2A_pp^{-s} + p^{1-2s}}{(1-p^{1-s})(1-p^{-s})} \prod_{p | 36} \frac{1}{(1-p^{1-s})(1-p^{-s})} \ , 
\end{equation}
where $A_p$ is an integer which can be fixed for each prime.
Comparing $L(s) = \prod\limits_{p \nmid 36} (1 - 2A_p + p^{2s-1})^{-1}$ gives us 
$(A_2, A_3, A_5, A_7, A_{11},\ldots) = (0,0,0,2,0,1,\ldots)$.

\section{Monsieur Mathieu}\label{s:moonshine}\setall
We have mentioned the sporadic group $M_{24}$ a few times throughout the text, which indeed was the original motivation for considering the cycle shapes.
Indeed, it was shown in \cite{FM} that the cycle shapes in fact encode the irreps of the sporadic group $M_{24}$.
Indeed, $M_{24}$, of order $2^{10} \cdot 3^3 \cdot 5 \cdot 7 \cdot 11 \cdot 23 = 244,823,040$, is a subgroup of the permutation group $\Sigma_{24}$ on 24 elements, generated by 2 elements which in standard cycle notation for permutations are (cf.~ \cite{lebryun})
\begin{align}
\nn M_{24} &:= \langle
(1,4,6)(2,21,14)(3,9,15)(5,18,10)(13,17,16)(19,24,23)
\ , \\
\label{genM24}
& \ \ 
(1, 4)(2, 7)(3, 17)(5, 13)(6, 9)(8, 15)(10, 19)(11,18)(12, 21)(14, 16)(20, 24)(22, 23)
\rangle \ .
\end{align}

Now, for permutation groups, cycle shapes are invariant under conjugation; therefore, conjugacy classes can be labeled thereby.
For the full $\Sigma_{24}$, there are of course the entire $\pi(24) = 1575$ number of conjugacy classes.
Here, for $M_{24}$, there are 26 conjugacy classes with 21 distinct cycle shapes, all of which appear in our list of 30, with the 9 exceptions being
$[6^3,2^3]; \ [9^2,3^3]; \ [8^2,4^2]; \ [6^4]; \ [22,2]; \ [20,4]; \ [18,6]; \ [16,8]; \ [12^2]$.

Consequently, the so-called ``multiplicative Moonshine phenomenon'' \cite{mason,FM} is the remarkable fact that the coefficients $a_p$ and $b_p$ in the L-function can all be expressed as virtual characters of $M_{24}$, that is, as simple $\IZ$-linear combinations of the entries of the (rational) character table of $M_{24}$.
The correspondence is precise in that the q-expansion of the eta-product of a particular cycle shape encodes the conjugacy class associated to that shape and is thus a McKay-Thompson series affiliated thereto.
For example, $[1^{24}]$, which is $\eta(z)^{24} = \Delta(z)$, should correspond to the class of the identity whence the dimensions of the irreps:
\begin{align}
\nn
\dim(\mbox{Irrep}_{M_{24}}) &= \{
1, 23, 45, 45, 231, 231, 252, 253, 483, 770, 770, 990, 990, 1035, 1035,\\
\nn
& \ 1035, 1265, 1771, 2024, 2277, 3312, 3520, 5313, 5544, 5796, 10395
\} \\
\end{align}
Indeed, the q-expansion of $\Delta(z)$ gives the Ramanujan tau-function, whose first values are
\begin{align}
\nn
\tau(n) &= \{
1, -24, 252, -1472, 4830, -6048, -16744, 84480, -113643, -115920, \\
\label{tau-coef}
&
\qquad \quad 534612, -370944, -577738, 401856, 1217160 \ldots
\} \ ,
\end{align}
and we have such simple linear combinations as
\begin{equation}
-24 = -1-23, \qquad
252 = 252, \qquad
-1472 =  1 + 23 - 231 - 1265, \ \ldots 
\end{equation}
expressing the $\tau$-coefficients in terms of the dimensions of irreps.

Now, in \cite{Eguchi:2010ej}, it was noticed that the elliptic genus of a K3 surface encodes the irreps of $M_{24}$ and thus began Mathieu Moonshine from the point of view of conformal field theory (cf.~\cite{GPRV,Govindarajan:2011em,umbral,Taormina:2013jza,Gannon:2012ck,Cheng:2010pq,Gaberdiel:2010ch,Gaberdiel:2010ca,Eguchi:2010fg,Gaberdiel:2012um,Cheng:2013kpa}).
In a recent work of \cite{umbral,Cheng:2013kpa}, this was realized as part of a web of string compactifications so that the elliptic genus corresponds to the partition function of $\cN=2$ type II string compactification on K3, and, by duality, the heterotic string on $K3 \times T^2$, much like our situation.
However, our eta-products are the generating functions of particular BPS spectra and differs from the elliptic genera as well as prepotential considerations of \cite{Eguchi:2010ej,Cheng:2013kpa}.
It would certainly be interesting to clarify the relations further, especially the role of multiplicativity in the conformal field theory.

For now, let us turn to a brief comparative study.
Recalling the theta-functions from \eqref{theta}, the statement of \cite{Eguchi:2010ej,Cheng:2013kpa} is that
\begin{align}
\nn
{\cal Z}_{K3}^{elliptic}(q,y) & = 8 \left[
    \left( \frac{\theta_2(q,y)}{\theta_2(q,1)} \right)^2 +
    \left( \frac{\theta_3(q,y)}{\theta_3(q,1)} \right)^2 +
    \left( \frac{\theta_4(q,y)}{\theta_4(q,1)} \right)^2
    \right] \\
&= -\frac{24 i y^{\frac12} \theta_1(q,y)}{\eta(q)^3} 
  \sum\limits_{n=-\infty}^\infty
  \frac{(-1)^n q^{\frac12n(n+1)}y^n}{1 - q^n y} + 
  \sum\limits_{n=0}^\infty A_n q^{n-\frac18} \frac{\theta_1(q,y)^2}{\eta(q)^3} \ ;
\end{align}
so that we have the coefficients $A_n = \{ −2, 90, 462, 1540, 4554, 11592, \ldots \}$.
In the original normalization of \cite{Eguchi:2010ej}, we halve these:
\begin{equation}\label{An-coef}
\tilde{A}_n = \{-1, 45, 231, 770, 2277, 5796, \ldots \}
\end{equation}
so that the simple combinations of the irreps of $M_{24}$ are even more apparent: these first few appear already in the irreps.

These two versions of Mathieu moonshine, multiplicative and elliptic, seem to extract different irreps as basis elements, as can be seen from \eqref{tau-coef} and \eqref{An-coef}.
In some sense, the two moonshine phenomena are complementary to each other.
There is, however, a relation between the Dedekind eta-function and the Jacobi theta-function: $\eta(q) = \frac{1}{\sqrt{3}} \theta_2(\frac{\pi}{6}, q^{\frac16})$, whence $\Delta(q) = \frac{1}{3^{12}} \theta_2(\frac{\pi}{6}, q^{\frac16})^{24}$, and thus at least part of the elliptic genus can be expressed in terms of the discriminant.
The reader is referred to the nice discussions in \cite{Cheng:2010pq} for the relations between how $M_{24}$ is encoded in these different aspects.

In our present context of K3 surfaces, as mentioned earlier, it is a classical result that any symmetry preserving the holomorphic 2-form on a K3 surface \cite{mukai} is a subgroup of $M_{24}$.
The essential reason for this is the fact that the homology lattice $H_*(K3, \IZ)$ is an even self-dual lattice of dimension 24 while $M_{24}$ is a natural (subgroup of) the automorphism group of such dimension 24 lattices.

\begin{figure}[!h!t!b]
\centerline{
\includegraphics[trim=0mm 0mm 0mm 0mm, clip, width=5.0in]{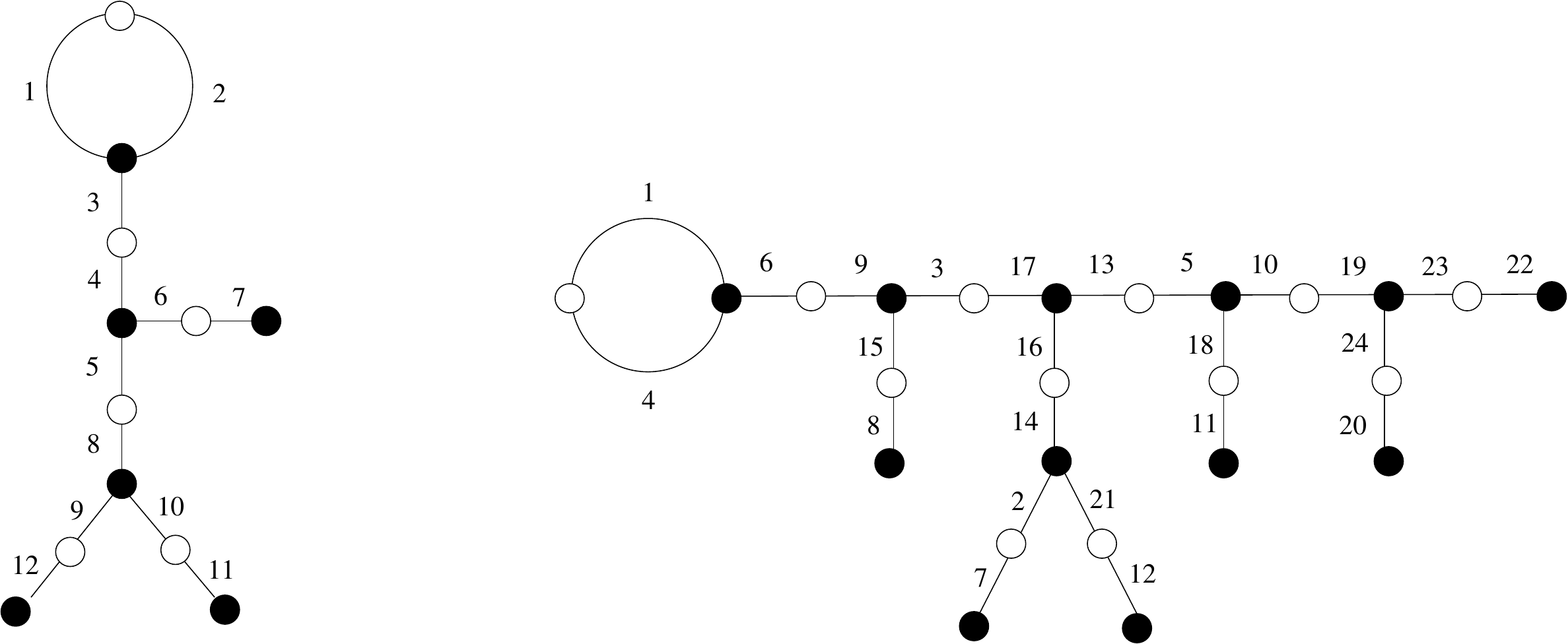}
}
\caption{{\sf {\small
{\it Monsieur Mathieu et son chien}:
The clean dessins d'enfants of $M_{12}$ on the left and that of $M_{24}$ on the right.
We label the 12 (respectively 24) edges corresponding to the elements of the set on which the permutation group $\Sigma_{12}$ (respectively $\Sigma_{24}$) acts.
}}
\label{f:M12M24}}
\end{figure}

Moreover, the family of Mathieu groups are constructible as dessins d'enfants, in suggestive figures which has been affectionately called ``Monsieur Mathieu'' \cite{lebryun}.
The subgroup $M_{12}$ of $M_{24}$ and of order 95040, itself one of the Mathieu family of 5 sporadic groups, generated by $s = (1,2)(3,4)(5,8)(7,6)(9,12)(11,10)$ and $t = (1,2,3)(4,5,6)(8,9,10)$, affords a particularly picturesque dessin; we present both of these dessins in Figure \ref{f:M12M24}.
Now, the generators of course have a degree of freedom in their choice and subsequently there are many ways to draw them \cite{zvonkin}, and the reader is referred to the classification results of \cite{HZ}
We adhere to these above two sets of generators of $M_{12}$ and $M_{24}$.

We note that both are {\em clean} dessins in the sense that the valency of one colour (here chosen as white) is always 2; we have also labeled the edges explicitly. As we read (counterclockwise) around each node, we have two sets of cycles, one for the black and the other for the white, these are precisely the generators of groups in cycle notation.
Indeed, we can form a so-called ``permutation triple'' by setting $\sigma_0$ and $\sigma_1$ as the black and white cycle generators in \eqref{genM24} and $\sigma_\infty$ such that $\sigma_0 \sigma_1 \sigma_\infty$ equals the identity permutation in $\Sigma_{24}$ (cf.~e.g.\cite{He:2012js} for rudiments, especially in the context of gauge theories).
Thus we have
\begin{align}
\nn
\sigma_0 &= (1,4,6)(2,21,14)(3,9,15)(5,18,10)(13,17,16)(19,24,23) \ ; \\
\nn
\sigma_1 &= (1, 4)(2, 7)(3, 17)(5, 13)(6, 9)(8, 15)(10, 19)(11,18)(12, 21)(14, 16)(20, 24)(22, 23) \ ; \\
\label{cartoM24}
\sigma_\infty &= (2,7,14,17,15,8,9,4,6,3,13,10,23,22,24,20,19,18,11,5,16,21,12)
\ .
\end{align}
The fact that $\sigma_\infty$ has only a cycle of length 23 and thus, trivially, a cycle of length 1, corresponds to the fact that the dessin for $M_{24}$ has only one visible face (a 2-gon).
\comment{
Indeed, while $M_{23}$ has been associated to K3 surfaces by \cite{mukai}, the connections between $M_{24}$ and K3 is only through Mathieu Moonshine \cite{GPRV,Govindarajan:2011em,umbral,Taormina:2013jza,Gannon:2012ck,Cheng:2010pq,Gaberdiel:2010ch,Gaberdiel:2010ca,Eguchi:2010fg,Gaberdiel:2012um,Cheng:2013kpa}.
However, the cartographic group in \eqref{carto24} corresponds to the ramification (passport) data ${\footnotesize \left\{ \begin{array}{c} 3^6,1^6 \\ 2^{12} \\ 23, 1\end{array}\right\}}$, which is one of our partitions in Table \ref{t:the30}, and thence, an elliptic K3 surface with cusp widths $[23,1]$.
}
The cartographic group in \eqref{carto24} corresponds to the ramification (passport) data ${\footnotesize \left\{ \begin{array}{c} 3^6,1^6 \\ 2^{12} \\ 23, 1\end{array}\right\}}$. Thus the pre-images of 0 are not tri-valent, whereby violating the conditions of \eqref{JmapBelyi}, and we do not actually have an elliptically fibred surface here.
What we do have is a trivalent clean dessin with six ``spikes'' \cite{mckaysebbar}, coming from the $1^6$ uni-valent nodes.
In principle we should be able to find a corresponding modular subgroup using the methods of \cite{He:2012jn}, though the group is expected to be neither genus 0 nor congruence.

\section{A Plethystic Outlook}\label{s:pleth}\setall
As a parting digression, let us see an alternative physical interpretation of the eta-products.
In \cite{Benvenuti:2006qr,Feng:2007ur} we proposed the Plethystic programme to study gauge theories, especially those with supersymmetry. The methods are very much in the spirit of the super-conformal index which has been introduced in \cite{Witten:2000nv} and extensively studied by \cite{Romelsberger:2005eg,Bhattacharya:2008zy,Gadde:2011uv}  et al.
Briefly, the programme follows the following steps:
\begin{itemize}
\item Find the vacuum geometry $\cM$ of the theory, which is the algebraic variety parametrized by the vacuum expectation values of the scalars. Compute the Hilbert series 
\begin{equation}
f(t) = \sum\limits_{n=0}^\infty a_n t^n \ , \qquad a_n \in \IZ_{\ge 0}
\end{equation}
of $\cM$ with respect to some appropriate grading dictated by the natural charges in the system. This is the generating function for counting the basic single-trace invariants;
\item To find the multi-trace objects, i.e., the unordered products of the single-traces, we take the plethystic exponential (sometimes know as the Euler transform)
\begin{equation}
g(t) =  PE[f(t)] := \exp\left( \sum_{n=1}^\infty \frac{f(t^n) -
  f(0)}{n} \right) = \frac{1}{\prod\limits_{n=1}^\infty (1-t^n)^{a_n}}  \ ;
  \end{equation}
\item 
There is an analytic inverse function to $PE$, which is the plethystic logarithm, given by
\begin{equation}
f(t) = PE^{-1}(g(t)) = \sum_{k=1}^\infty
\frac{\mu(k)}{k} \log (g(t^k)) \ , 
\end{equation}
where
\[
\mu(k) := \left\{\begin{array}{lcl}
0 & & k \mbox{ has repeated prime factors}\\
1 & & k = 1\\
(-1)^n & & k \mbox{ is a product of $n$ distinct primes} \end{array}\right.
\] 
where $\mu(k)$ is the M\"obius function. The plethystic logarithm
of the Hilbert series gives the geometry of $\cM$, i.e., 
\[
PE^{-1}[f(t)] = \mbox{ defining equation of $\cM$}.
\] 
In particular, if $\cM$ were complete-intersection variety, then $PE^{-1}[f(t)]$ is polynomial;
\item The Hilbert series of the $N$-th symmetric product is given by
\begin{equation}
g_N(t; \cM) = f(t;\sym^N(\cM)), \qquad \sym^N(\cM) := \cM^N/S_N \ ,
\end{equation}
where the ``grand-canonical'' partition function is given by the fugacity-inserted plethystic exponential of the Hilbert series:
\begin{equation}
g(\nu ; t) = PE_\nu[f(t)] := \prod\limits_{n=0}^{\infty} \frac{1}{(1
- \nu  \, t^n)^{a_n}} = \sum\limits_{N=0}^\infty g_N(t) \nu^N  \ .
\end{equation}
In the gauge theory, this is considered to be at finite $N$.
\end{itemize}

We mentioned in \S\ref{s:boson} that the most natural manipulation to perform upon multiplicative series is to take the Dirichlet transform.
Indeed, the Riemann zeta function itself can be seen as the Dirichlet transform of the plethystic logarithm of $\varphi(q)$ in the following way:
$PE^{-1}[\varphi(q)] = (1-q)^{-1} = \sum\limits_{n=0}^\infty a_n t^n$ with $a_n = 1$ and whence $\sum\limits_{n=1}^\infty n^{-s} = \zeta(s)$), 

Inspired by \eqref{Gbosonic} and given now the wealth of multiplicative function constructed from eta-products, from the aper\c{c}u of the plethystic programme, we need to compute
\begin{equation}
PE^{-1}[q / F(q)] \mbox{ where $F(q)$ is a multiplicative eta-product}
\ ,
\end{equation}
treating $q$ purely as a formal ``dummy'' variable in the generating function.
Note that the $q$ in the numerator serves to cancel the product over $q^{1/24}$ which always yields $q$ in the denominator for our partitions.
In the ensuing, we will projectivize our varieties for convenience and geometrically interpret our Hilbert series as that of projective varieties, possibly with
 weights.

The simplest case of $[1^{24}] = \eta(z)^{24}$ in \eqref{eta24} gives
\begin{equation}
PE^{-1}[q / \eta(q)^{24}] = 24 PE^{-1}[\varphi(q)] = \frac{24}{1-q} \ .
\end{equation}
Using the standard method of interpretation \cite{BS}, this is simply 24 distinct points in general position on $\IP^1$.
Note that this is, of course, geometrically different from $\IP^{24}$, which would have the Hilbert series $(1-q)^{-24}$.

Let us now move onto a more non-trivial one, say $[2^8,1^8] = \eta(2z)^8 \eta(z)^8$, whereby
\begin{equation}
PE^{-1}[\frac{q}{\eta(q^2)^{8}\eta(q)^8}] = 
PE^{-1}[\varphi(q^2)^8 \varphi(q)^8] =
\frac{8}{1 - q^2} + \frac{8}{1 - q}  \ .
\end{equation}
Similar to the above, this is the Hilbert series of 16 points in weighted $\IP^1$ with weights $[1:2]$ on the projective coordinates, though not in general position so that linear relations exist amongst them.
In general, because our cycle shape is always of the form $[n_1^{a_1}, \ldots, n_t^{a_t}]$ with one of the $n_i$, say the first, dividing all other $n_i$, or in the simplest case, with just a single $[n_1^{a_1}]$.
Therefore, after taking the plethystic logarithm, we will always have the Hilbert series of the form $\sum\limits_{i=1}^t \frac{a_i}{1 - q^{n_i}}$.
The leading fractional contribution $\frac{a_1}{1-q^{n_1}}$ has a denominator which divides all others, thus allowing the remaining fractions to be combined to yield a final answer $PE^{-1}[q / F(q)] = \frac{a_1 + f(q)}{1 - q^{n_1}}$ for some polynomial $f(q)$.
This geometrically represents points in weighted-$\IP^1$ with weights $[1:n_1]$ which are not necessarily in general position.

Indeed, the physical origins between our main discussion on the eta-products as partition functions of certain BPS states in type IIA on $K3 \times T^2$ and this abovementioned view-point on the geometry encoded by the Hilbert series, generically arising from type IIB on Calabi-Yau spaces, are different, though tantalizingly similar.
It would be fascinating to see whether there might be some mirror-type of transformation which maps one to the other.

\comment{
In Macaulay2:
http://www.math.uiuc.edu/Macaulay2/doc/Macaulay2-1.6/share/doc/Macaulay2/Macaulay2Doc/html/_hilbert__Series_lp__Projective__Hilbert__Polynomial_rp.html

To find the Hilbert series of Projective varieties, can't use hilbertSeries
directly, have to use:
 hilbertSeries hilbertPolynomial ideal ( ... )
}

\section{Conclusions and Prospects}\label{s:conc}
Having indulged ourselves with two parallel strands of thought, let us pause here for a brief reflection.
Motivated by the relation of the multiplicative structure of the (reciprocal) generating function for the oscillator modes of the bosonic string, we have commenced with the full list of products of Dedekind eta functions which are multiplicative and have subsequently delved into the compactification of the heterotic string on appropriate six-tori whose generating functions of BPS states are known to be precisely this list, 30 in number.
These products further possess modular properties and are, in fact, certain modular forms of weight $k$ which is equal to half the number of terms in the product.
By string duality, the type IIB realization of this compactification is that of K3 surfaces.
The multiplicative constraint singles out special K3 surfaces which admit Nikulin involutions, falling under 14 classes.
Indeed, all these K3 surfaces are in the list of the 30 and correspond to the situation where $k \ge 3$. In the algebraic realization of elliptic fibration over $\IP^1$, these K3 surfaces are all semi-stable with $2k$ Kodaira type-I fibres.
Central to the above are particular partitions of the number 24.

Along another vein relating to semi-stable K3 surfaces, there is a partitioning problem of 24 of which there is a classification totaling 112 which are extremal in the sense of possessing 6 singular type-I fibres.
These all correspond to subgroups of the modular group by having the dessins d'enfants corresponding to their J-maps identifiable with the Schreier coset graph of the modular subgroup.
Equivalently, the subgroup is also the cartographic group of the dessin.
Of these 112, nine are congruence and genus zero and have been investigated in the context of $\cN=2$ gauge theories in four dimensions.

We have inter-woven the co-extending skeins by showing that the two sets of K3 surfaces and partitioning, whenever intersecting, are in fact the same geometries by finding the explicit Weierstra\ss\ models.
That multiplicativity and modularity should engender the same geometries, both affording interpretations as string compactifications is fascinating and merit further investigations.
In \cite{He:2012kw}, the proposal was made that a class of Gaiotto $\cN=2$ theories in four dimensions should be obtainable for every K3 surface who J-invariant is Belyi, and indeed for any connected finite index subgroup of the modular group.
It would be interesting to see how these gauge theories relate, when possible, to the gauge theories obtainable from the compactification on $T^2$ times a Nikulin K3 surface.

Furthermore, we have discussed how the eta-products encode the characters of the sporadic group $M_{24}$.
This is particularly relevant given the recent explosion of activity on Mathieu moonshine, especially in the interpretations of elliptic genera of K3 surfaces and partition functions of the dual heterotic compactification.
It is interesting how our version is complementary to the ones obtained in the literature; this is certainly worthy of further investigation.

Another fascinating direction to take is to follow the works of Yau and Zaslow \cite{Yau:1995mv}.
There, the authors realized that the number $n_d$ of degree $d$ rational curves on a K3 surface obeys, in fact, the generating function
\begin{equation}
\sum\limits_{d=1}^\infty q^d = q \eta(q)^{-24} \ ,
\end{equation}
which is precisely the counting function discussed in \eqref{Gbosonic} that initiated our quest.
That the Riemann Hypothesis could be translated, via a theorem of Lagarias \cite{lagarias}, to a statement on rational curves on K3 surfaces using the above fact, was discussed in \cite{He:2010jh}.
It is therefore natural to enquire whether all our eta-products afford interpretations as Gromov-Witten invariants.
On these and many more lines of enquiry we shall pursue.

\section*{Acknowledgements}
We are indebted to helpful comments from and discussions with Miranda Cheng, Llyod Kilford, Viacheslav Nikulin, Simon Norton, Wissam Raji.
YHH would like to thank the Science and
Technology Facilities Council, UK, for grant ST/J00037X/1, 
the Chinese Ministry of Education, for a Chang-Jiang Chair Professorship at NanKai University as well as the City of Tian-Jin for a Qian-Ren Scholarlship, the US NSF for grant CCF-1048082, as well as City University, London and Merton College, Oxford, for their enduring support.
Moreover, he is indebted to the kind hospitality to McGill University and to the Perimeter Institute where the final stages of this work were completed.
JM is grateful to the NSERC of Canada.

~\\
~\\
~\\


\appendix
\section{Further Salient Features of Eta}
In this appendix, we collect some further properties of the eta-function, ranging from standard modularity arguments to combinatorial interpretations of their products and quotients.
\subsection{Modularity}\label{ap:eta-modular}
It is a standard fact that the Dedekind eta function
\begin{equation}
\eta(q) = q^{\frac{1}{24}} \prod\limits_{n=1}^\infty (1-q^n) \ ,
\qquad q = e^{2 \pi i z} \ , \ z \in \cH
\end{equation}
is a modular form of weight $\frac12$ on the upper half plane $\cH$ 
(cf.~e.g.,\S III.2 of \cite{koblitz} or a classic of Siegel from the perspective of residues in \cite{siegel}).

It is illustrative to show the workings of the action by the modular group.
First, under $z \mapsto z + 1$, we clearly have that $\eta(z+1) = \exp(\frac{\pi i}{12}) \eta(z)$.
Next, for the transformation $z \mapsto -1/z$, consider the logarithmic derivative.
To fully appreciate the prefactor, let us define
\begin{equation}
\tilde{\eta}(z) := \prod\limits_{n=1}^\infty (1-q^n) \ .
\end{equation}
Whence,
\begin{equation}
\frac{\tilde{\eta}'(z)}{\tilde{\eta}(z)}  
= \sum_{n=1}^\infty 
   \frac{(-2\pi i n) e^{2 \pi i n z}}{1 - e^{2 \pi i n z}} 
= - 2 \pi i  \sum_{n=1}^\infty \sigma_1(n) q^n \ .
\end{equation}
where $\sigma_k(n) := \sum\limits_{d | n} d^k$ is the divisor sum function and where we have used the standard Lambert sum:
\begin{equation}
\sum_{n=1}^\infty \frac{n^k q^n}{1 - q^n} = 
     \sum_{n=1}^\infty \sigma_k(n) q^n \ .
\end{equation}

Finally, we recall that the normalized Eisenstein series is itself a sum over divisor functions (cf.~\cite{koblitz,serre} and adhering to the conventions of the latter)
\begin{equation}
E_k(z) = \frac12 \sum_{{\tiny \begin{array}{c} m,n \in \IZ \\ \gcd(m,n) = 1 \end{array}}}
    (m z + n)^{-k} = 1 - \frac{2k}{B_k} \sum_{n=1}^\infty \sigma_{k-1}(n) q^n
, \quad k \in 2 \IZ_{>0}
\end{equation}
where $B_k$ is the $k$-th Bernoulli number and that
\begin{equation}\label{Ek}
z^{-2} E_2(-\frac{1}{z}) = E_2(z) + \frac{12}{2\pi i z} \ ,
\quad
z^{-k} E_k(-\frac{1}{z}) = E_k(z) \ , \qquad k \in 2 \IZ, k > 2 \ ,
\end{equation}
(so that indeed, for even $k>2$ we have modular forms of weight $k$ and for $k=2$, we have the extraneous term $\frac{12}{2\pi i z}$).
Hence, the logarithmic derivative is
\begin{equation}\label{eta-E}
\frac{\tilde{\eta}'(z)}{\tilde{\eta}(z)} = \frac{2 \pi i B_2}{4}(1 - E_2(z)) \ .
\end{equation}

The inhomogeneity of the above, in relation to $E_2$, will inevitably ruin any nice modular behaviour.
This is why the Dedekind function has the extra power of $q^{\frac{1}{24}}$ so as to modify \eqref{eta-E} to (note the reciprocal removes the minus sign in front of $E_2(z)$)
\begin{equation}
\frac{\eta'(z)}{\eta(z)} = \frac{2 \pi i}{24} E_2(z)
,
\end{equation}
so that
\begin{equation}
z^{-2} \frac{\eta'(-1/z)}{\eta(-1/z)} = 
\frac{2 \pi i}{24} E_2(z) + \frac{2 \pi i}{24} \frac{12}{2\pi i z} = 
\frac{1}{2z} + \frac{\eta'(z)}{\eta(z)}
\end{equation}
on using \eqref{Ek}.
Hence, $\eta(-1/z) = \eta(z) \sqrt{z C}$ for some constant $C$ upon integration, which can be fixed to be $-i$ by substituting $z=i$.

In general, under ${\tiny \left( \begin{matrix}
a & b \\ c & d
\end{matrix} \right)} \in SL(2;\IZ)$, we have that
\begin{align}
\nn
&\eta(\frac{a z + b}{c z + d}) = 
  (c z + d )^{\frac12} \  \chi_{a,b,c,d} \ \eta(z) \ ,
\\
&
\qquad 
\chi_{a,b,c,d} = \left\{
\begin{array}{lcl}
\exp(\frac{b \pi i}{12})  & \ , & c = 0, d = 1 \\
\exp(\pi i \left( \frac{a+d}{12c} - \frac14 - \sum\limits_{n=1}^{c-1}
\frac{n}{c} \left( \frac{dn}{c} - \lfloor \frac{dn}{c} \rfloor - \frac12
\right) \right)) & \ , & c > 0 \ .
\end{array}
\right.
\end{align}

\subsection{Some Partition Identities}
We collect some interesting properties of the Euler function which encode various partitions, some are the explicit q-expansions of our eta-products; cf.~\cite{kac1,ono}:
\[
\begin{array}{rcl}
\mbox{Euler, 1748}&&\eta(q) = q^{\frac{1}{24}}\sum\limits_{k=-\infty}^\infty 
     (-1)^k q^{\frac{3k^2+k}{2}} \\
\mbox{Jacobi, 1828}&&\eta(q)^3 = q^{\frac{1}{8}}\sum\limits_{k=-\infty}^\infty 
     (4k+1) q^{2k^2+k} 
\ , \qquad
\eta(q^8)^3 = q \sum\limits_{k=0}^\infty (-1)^k (2k+1) q^{(2k+1)^2} 
\\
\mbox{Gauss, 1866}&& \frac{\eta(q)^2}{\eta(q^2)} = \sum\limits_{k=-\infty}^\infty 
     (-1)^k q^{k^2}
\ , \qquad
\frac{\eta(q^2)^2}{\eta(q)} = q^{\frac{1}{8}} \sum\limits_{k=-\infty}^\infty q^{2k^2+k} 
\\
\mbox{Gordon, 1961}&&\frac{\eta(q^2)^5}{\eta(q)^2}=
q^{\frac{1}{3}}\sum\limits_{k=-\infty}^\infty (-1)^k (3k+1)q^{3k^2+2k} 
\ , \qquad
\frac{\eta(q)^5}{\eta(q^2)^2}=q^{\frac{1}{24}}\sum\limits_{k=-\infty}^\infty 
     (6k+1)q^{\frac{3k^2+k}{2}} 
\\
\mbox{Macdonald, 1972}&&\frac{\eta(q^6)^5}{\eta(q^3)^2}=
\sum\limits_{k=1}^\infty (-1)^{k-1} \jacobi{k}{3} k q^{k^2} 
\\
\mbox{Kac, 1980}
&&
\eta(q^{12})^2 = \sum\limits_{k,m\in \IZ, k \ge 2|m|} (-1)^{k+m} 
q^{\frac{3(2k+1)^2-(6m+1)^2}{2}}
\\
&&\eta(q^{16})\eta(q^8) = \sum\limits_{k,m\in \IZ, k \ge 3|m|} (-1)^{k} 
q^{(2k+1)^2-32 m^2}
\\
&&
\eta(q^{20})\eta(q^4) = \sum\limits_{k,m\in \IZ_{\ge 0}, 2k \ge m} (-1)^{k} 
q^{\frac{5(2k+1)^2-(2m+1)^2}{4}}
\\
\end{array}
\]

\section{The $j$-function: Partition Properties}
One can write the $j$-invariant in terms of our Dedekind $\eta$-function:
\begin{equation}\label{j-eta}
j(q) = 64 \frac{(t+4)^3}{t^2} \ , \qquad
t := \frac{1}{64} \left( \frac{\eta(z)}{\eta(2z)} \right)^{24} \ ,
\end{equation}
wherein we could further write in terms of the partition $\pi_n$ of integers from \eqref{phi}:
\begin{equation}
q^{\frac{1}{24}} (\eta(z))^{-1} = \sum_{k=0}^\infty \pi_k q^k \ .
\end{equation}
Subsequently, we can substitute and expand $j(q)$ in terms of $\pi_n$ to find
\begin{dmath}\label{j-partition}
j(q) = 
\frac{1}{q}+
\frac{24 \left(32 \pi_0-\pi_1\right)}{\pi_0}
+
\frac{12}{\pi_0^2} \left(16384 \pi_0^2+2 \pi_1 \pi_0-2 \pi_2 \pi_0+25 \pi_1^2\right) q
+
\frac{8}{\pi_0^3}
   \left(2097152 \pi_0^3+589824 \pi_1 \pi_0^2-3 \pi_3 \pi_0^2-72 \pi_1^2 \pi_0+75 \pi_1 \pi_2 \pi_0-325 \pi_1^3\right) q^2
+
\frac{6}{\pi_0^4} \left(2925
   \pi_1^4+1200 \pi_0 \pi_1^3+9044014 \pi_0^2 \pi_1^2-1300 \pi_0 \pi_2 \pi
   _1^2+133431296 \pi_0^3 \pi_1-96 \pi_0^2 \pi_2 \pi_1+100 \pi_0^2 \pi_3 \pi
   _1+50 \pi_0^2 \pi_2^2+786436 \pi_0^3 \pi_2-4 \pi_0^3 \pi_4\right) q^3
+
\cO\left(q^4\right)
\end{dmath}

Substituting in the standard first values
\begin{equation}
(\pi_0, \pi_1, \pi_2, \pi_3 \, \ldots) = 
(1,1,2,3,5,7, \ldots)
\end{equation}
readily retrieves the famous coefficients 744, 196884, etc.
We note that the numerators are all homogeneous polynomials in the partitions $\pi_i(n)$, one naturally questions oneself what significance they carry.

Alternatively, we can use the Eisenstein series
\begin{eqnarray}
\nn
j(q) &=& 1728 \frac{g_2(q)^3}{g_2(q)^3 - 27 g_3(q)^2} =
1728 \frac{g_2(q)^3}{\Delta(q)}
\\
&=&\frac{(1 + 240 \sum\limits_{n \ge 1} \sigma_3(n) q^n)^3}{q\prod\limits_{n\ge 1} (1-q^n)^{24}} = \frac{1}{q} \varphi(q)^{24} (1 + 240 \sum\limits_{n \ge 1} \sigma_3(n) q^n)^3 \ .
\label{j-cube}
\end{eqnarray}
This gives us a positive combination in terms of the partition numbers $\pi_n$ and the divisor function $\sigma_3(n)$, which are themselves positive integers;
this is obviously a useful expansion for $j(q)$:
\begin{eqnarray}
\nn
&&\frac{1}{q} \pi_0^{24}+\\
\nn
&&\quad \left(720 \pi _0^{24} \sigma _3(1)+24 \pi _1 \pi_0^{23}\right)+ \\
\nn
&& q \left(172800 \pi _0^{24} \sigma _3(1){}^2+720 \pi _0^{24} \sigma
   _3(2)+17280 \pi _1 \pi _0^{23} \sigma _3(1)+24 \pi _2 \pi _0^{23}+276 \pi _1^2 \pi_0^{22}\right)+ \\ 
\nn
&& q^2 \left(13824000 \pi _0^{24} \sigma _3(1){}^3+345600 \pi _0^{24}
   \sigma _3(1) \sigma _3(2)+720 \pi _0^{24} \sigma _3(3)+ \right.
\\
\nn
&&
 \qquad 4147200 \pi _1 \pi _0^{23}
   \sigma _3(1){}^2+17280 \pi _2 \pi _0^{23} \sigma _3(1)+17280 \pi _1 \pi _0^{23}
   \sigma _3(2)+198720 \pi _1^2 \pi _0^{22} \sigma _3(1)+\\
\nn
&&
\qquad \quad
\left.
24 \pi _3 \pi _0^{23}+552
   \pi _1 \pi _2 \pi _0^{22}+2024 \pi _1^3 \pi _0^{21}\right)
+\cO(q^3) 
\label{j-part-sigma}
\end{eqnarray}
One could, of course, equate the two expansions \eqref{j-partition} and \eqref{j-part-sigma}, to obtain expressions for $\pi_n$ in terms of $\sigma_3(n)$ (for convenience, we have set $\sigma_3(0)=1$ as is the convention):
\begin{eqnarray}
\nn
\pi_1 &=& 16-15 \sigma _3(1), \\ 
\nn
\pi_2 &=& \frac{3825}{2} \sigma_3(1){}^2-\frac{12015 \sigma _3(1)}{2}-15 \sigma _3(2)+4232, \\
\nn
\pi _3 &=& 
   -\frac{631125}{2} \sigma _3(1){}^3+\frac{3690225}{2} \sigma _3(1){}^2+3825 \sigma
   _3(2) \sigma _3(1)-3102000 \sigma _3(1)-\\
\nn
&&\qquad - 6000 \sigma _3(2)-15 \sigma
   _3(3)+1592448
\\
\end{eqnarray}

Likewise, one could use the following expression, which is used to prove identities of the Ramanujan tau-function,
\begin{equation}
j(q) - 1728 = \varphi(q)^{24} \left( 1 - 504 \sum\limits_{n \ge 1} \sigma_5(n) q^n \right)^2 \ ,
\label{j-sq}
\end{equation}
to express all the partitions in terms of $\sigma_5$.

\subsection{q-Expansion of Roots of the $j$-function}
We have exploited the relation between the $j$-function and the eta-function in the above to obtain expressions of the famous $q$-coefficients of the former in terms of the partition numbers.
Here, we tabulate a few interesting but perhaps less known expansions for the various roots of the $j$-function. 
First, the famous $q$-expansion of the j-function is
\begin{dmath}
j(q) =  \frac{1}{q}+744+196884 q+21493760 q^2+864299970 q^3+20245856256 q^4+333202640600 q^5+
4252023300096 q^6+44656994071935 q^7+401490886656000 q^8+
3176440229784420
   q^9+22567393309593600 q^{10}+146211911499519294 q^{11}+874313719685775360
   q^{12}+4872010111798142520 q^{13}+25497827389410525184 q^{14} + \ldots
\end{dmath}

The $n$-roots of $j(q)$ afford integer $q$ expansions when $n$ is a divisor of 24.
In particular, we have the following:
\begin{dmath}
j(q)^{1/2} = 
\frac{1}{\sqrt{q}}+372 \sqrt{q}+29250 q^{3/2}-134120 q^{5/2}+54261375
   q^{7/2}-6139293372 q^{9/2}+854279148734 q^{11/2}-128813964933000
   q^{13/2}+20657907916144515 q^{15/2}-3469030105750871000
   q^{17/2}+603760629237519966018 q^{19/2}
+ \ldots
\end{dmath}

\begin{dmath}\label{cube}
j(q)^{1/3} = 
\frac{1}{q^{1/3}}+248 q^{2/3}+4124 q^{5/3}+34752 q^{8/3}+213126 q^{11/3}+1057504
   q^{14/3}+4530744 q^{17/3}+17333248 q^{20/3}+60655377 q^{23/3}+197230000
   q^{26/3}+603096260 q^{29/3}
+ \ldots
\end{dmath}

\begin{dmath}
j(q)^{1/4} = 
\frac{1}{q^{1/4}}+186 q^{3/4}-2673 q^{7/4}+430118 q^{11/4}-56443725
   q^{15/4}+8578591578 q^{19/4}-1411853283028 q^{23/4}+245405765574252
   q^{27/4}-44373155962556475 q^{31/4}+8266332741845429800
   q^{35/4}-1576306833508315403544 q^{39/4}
+ \ldots
\end{dmath}

\begin{dmath}
j(q)^{1/6} = 
\frac{1}{q^{1/6}}+124 q^{5/6}-5626 q^{11/6}+715000 q^{17/6}-104379375
   q^{23/6}+16966161252 q^{29/6}-2946652593626 q^{35/6}+535467806605000
   q^{41/6}-100554207738307500 q^{47/6}+19359037551684042500
   q^{53/6}-3800593180746056684372 q^{59/6}
+ \ldots
\end{dmath}

\begin{dmath}
j(q)^{1/8} = 
\frac{1}{q^{1/8}}+93 q^{7/8}-5661 q^{15/8}+741532 q^{23/8}-113207799
   q^{31/8}+19015433748 q^{39/8}-3390166183729 q^{47/8}+629581913929419
   q^{55/8}-120437982238038210 q^{63/8}+23564574046009042869
   q^{71/8}-4692899968498921291530 q^{79/8}
+ \ldots
\end{dmath}

\begin{dmath}
j(q)^{1/12} = 
\frac{1}{q^{1/12}}+62 q^{11/12}-4735 q^{23/12}+651070 q^{35/12}-103766140
   q^{47/12}+17999397756 q^{59/12}-3292567703035 q^{71/12}+624659270035130
   q^{83/12}-121698860487451255 q^{95/12}+24194029851560118900
   q^{107/12}-4886913657541566648179 q^{119/12}
+ \ldots
\end{dmath}

\begin{dmath}
j(q)^{1/24} = 
\frac{1}{q^{1/24}}+31 q^{23/24}-2848 q^{47/24}+413823 q^{71/24}-68767135
   q^{95/24}+12310047967 q^{119/24}-2309368876639 q^{143/24}+447436508910495
   q^{167/24}-88755684988520798 q^{191/24}+17924937024841839390
   q^{215/24}-3671642907594608226078 q^{239/24}
+ \ldots
\end{dmath}

Of these, the most discussed one is \eqref{cube} which has all positive integer coefficients and corresponds to the McKay-Thompson series for the Class 3C for the Monster Group.
Remarkably, it also encodes the irreducible dimensions of $E_8$.
These are discussed in \cite{kac1,kac2}.

Using \eqref{j-eta} and being mindful of the product formulae for the eta function, we can write
\begin{equation}
j(z) = 2^8 \cdot 3 + \frac{\eta(z)^{24}}{\eta(2z)^{24}}
+ 2^{16} \cdot 3 \frac{\eta(2z)^{24}}{\eta(z)^{24}} +
2^{24} \frac{\eta(2z)^{48}}{\eta(z)^{48}} \ .
\end{equation}
In general for the various roots wherein $d=1,2,3,4,6,8,12,24$,
\begin{equation}
j(z) = \frac{(\eta(z)^{24} + 2^{8} \eta(2z)^{24})^{3/d}}{\eta(z)^{48/d}\eta(2z)^{24/d}} =
\frac{(\varphi(q^2)^{24} + 2^{8} q \varphi(q)^{24})^{3/d}}{q^{1/d}\varphi(q)^{24/d}\varphi(q^2)^{48/d}} \ .
\end{equation}

We can use expressions \eqref{j-cube} and \eqref{j-sq} to simplify two of the roots.
For the cubic root, we see why immediately all the coefficients are positive:
\begin{equation}
j(q)^{1/3} = \frac{1}{q^{1/3}} \varphi(q)^{8} (1 + 240 \sum\limits_{n \ge 1} \sigma_3(n) q^n) =  \frac{1}{q^{1/3}} \left( \sum\limits_{n \ge 0} \pi_n q^n\right)^8
(1 + 240 \sum\limits_{n \ge 1} \sigma_3(n) q^n)
.
\end{equation}
From this, we can see how to write the McKay-Thompson series coefficients for class 3C of the Monster in terms of polynomials in $\pi_n$ and $\sigma_3(n)$ with positive coefficients.

For the square root, we see that
\begin{equation}
(j(q)-12^3)^{1/2} = \varphi(z)^{12} \left( 1 - 504 \sum\limits_{n \ge 1} \sigma_5(n) q^n \right) = \left( \sum\limits_{n \ge 0} \pi_n q^n\right)^{12} 
\left( 1 - 504 \sum\limits_{n \ge 1} \sigma_5(n) q^n \right) \ .
\end{equation}
Note that we need the shift of the constant term by $12^3 = 1728$ in order to get the perfect square.
Upon expansion, we obtain
\begin{dmath}
(j(q)-12^3)^{1/2} = 
\frac{1}{\sqrt{q}}-492 \sqrt{q}-22590 q^{3/2}-367400 q^{5/2}-3764865 q^{7/2}-28951452
   q^{9/2}-182474434 q^{11/2}-990473160 q^{13/2}-4780921725 q^{15/2}-20974230680
   q^{17/2}-84963769662 q^{19/2}+O\left(q^{21/2}\right) \ .
\end{dmath}
The coefficients here are all negative and the magnitudes thereof are precisely the McKay-Thompson series for Class 2a of the Monster \cite{FMN}.
Thus of all these 7 roots of the $j$-function, $n=2,3$ have been given nice interpretation, the integers in the remaining 5 are still elusive.

\newpage


\end{document}